\documentclass{article}
\usepackage{amsmath,amsfonts,amsthm}
\usepackage{amstext}
\usepackage{amsgen}
\usepackage{amsbsy}
\usepackage{amsopn}
\usepackage{amssymb}
\usepackage{graphicx}
\usepackage{epsfig}
\theoremstyle{definition}

\theoremstyle{remark}

\numberwithin{equation}{section}
\newtheorem{theorem}{Theorem}

\newtheorem{proposition}[theorem]{Proposition}
\newtheorem{remark}[theorem]{Remark}

\begin{document}

\title{Fully Algebraic Description of the Static Level Sets
for the System of Two Particles under a Van der Waals Potential}
\author{C.~Barr\'on-Romero$^{1,2}$, A.~Cueto-Hern{\'a}ndez$^{2}$, and F.~Monroy-P{\'e}rez$^{2}$
 \\
$^1$ cbarron@correo.azc.uam.mx\\ \\
$^2$ UAM-Azcapotzalco \\
Department of Basic Sciences \\
Av. San Pablo No. 180, Col. Reynosa Tamaulipas,\\
 C.P. 02200, MEXICO }

% \thanks{}%
% \subjclass{}%

\date{}
%\dedicatory{}%
%\commby{}%

\maketitle

% ----------------------------------------------------------------
\begin{abstract}

We study the equipotential surfaces around of a two particle system in
3-d under a pairwise good potential as the one of Van der Waals. The
level sets are completely determined by the solutions of polynomials of
at most fourth degree that can be solved by standard algebraic methods.
The distribution of real positive roots determines the level sets and
provides a complete description of the map for the equipotential zones.
Our methods can be generalized to a family of polynomials with degree
multiple of 2, 3, or 4. Numerical simulations of 2-d and 3-d pictures
depicting the true orbits and equipotential zones are provided.


\end{abstract}
% ----------------------------------------------------------------
\section{Introduction}

The study of clusters of massless non-interactive particles under a
potential fields has been approached, roughly speaking, under two view
points: the one that considers the dynamics given by equations of
Newton associated to the potential, and the one that considers static
particles. The latter is related among others, to theoretical and
experimental research in crystal formation, and plasticity, whereas the
former is connected with research in control theory. Another line of
research involving pairwise good potentials and control of chaos
concerns to the dynamical analysis and control of micro cantilevers as
single mode approximation for tackling interactions under Van der Waals
potential, see for instance \cite{cantiliver:MR1725890}.

In the study of the N-body problem, several pairwise good potential
have been utilized for describing relative equilibria and central
configurations, see for instance ~\cite{phyAstr:Corbera2004}, the
methods there include classical dynamical systems as well as non-linear
geometric control techniques \cite{Bonnard:MR2737879}.

%In finite elastoplasticity, Lie groups and algebras have been recently
%applied, for instance in~\cite{Mielke:MR1919826}, A. Mielke associates
%to the plastic tensor a certain element of a Lie group, and to the
%plastic dissipation a Finsler metric of the same Lie group.
%Following this
%approach, D. Mittenhuber \cite{AtXiv:Mittenhuber2002}, has solved
%the problem of 4-slip system in the plane with orthogonal slip
%directions by computing the associated dissipation metric as the
%solution of  an optimal control problem.

We are interested in clusters of non-interactive particles under
pairwise good potentials, our research points to the optimal control
problems of optimal path planning, collision free navigation and
crystal formation on equipotential zones of clusters.

 In this paper, we
restrict ourselves to the study of the equipotential zones for a system
of two static non interactive particles.
%, as the preliminary step for continuing the control theoretic approach
%of the aforementioned problems.
%To the best of our knowledge, most of the reported research on clusters
%of particles under a well pair potential is focused on numerical
%methods for the problem of search of optimal clusters, see for
%instance~\cite{jgo:Maranas1994, jpc:Wolf1998,cpc:Romero1999A,
%ape:Solovyov2003, jpca:Xiang2004A}, and the references therein. In this
%paper we take a different approach from numerics.

In the literature, numerical methods of mathematical software is
commonly used to generate graphics of  levels set or orbits. However,
there are not algebraic methods and techniques for determining the
shape of the surfaces or curves or dots of the equipotential energy
even for small clusters.
 The novelty of our approach is
to solve the implicit function problem of the determination of the
level sets using the algebraic methods for solving polynomials by
radicals. With our methods, the exact solution of an orbit corresponds
to roots of a polynomial with degree $n  \leq 4$ as formul\ae \
involving its coefficients, arithmetic operations, and radicals.

%We also show that a numerical approximation could give a ``good"
%numerical approximation to the solution that is visually  similar to
%the true orbit.

% As we mentioned before, this is a first step of a general research program that
% pretends to tackle optimal control problems on clusters under pairwise good potentials.

 The paper is organized as follows, in section~\ref{sc:notation}, we depict the
 notation of the problem,
 the general description of our
 approach, and our main result. Section~\ref{sc:Polynomial} contains
  the description of the algebraic procedures. In section~\ref{sc:Results} we present
 a comparison of numerical and algebraic results.
 At the end in section~\ref{sc:conclusions and future work},
 we derive some conclusions.
 % and generally describe some of the future lines of research.

\section{Notation and Problems}~\label{sc:notation}

Let  be $R^+ =[0,\infty)$. For a given  $d>0$, a smooth function
$P:\mathbb{R}^+ \rightarrow \mathbb{R}$ is well pairwise good potential
if satisfies the following conditions:

\begin{enumerate}
    \item Infinite rejection for avoiding, destroying or collapsing particles, $\lim_{d \to 0}
    P(d)=\infty$.
    \item A negative basin around a minimum distance $d^\star = \arg \min_{d \in (0,\infty)} P(d)$.
    \item Asymptotic attraction, $\lim_{d\to\infty} P(d)=0^{-}$.
\end{enumerate}

An elementary model for the Van der Waals that complies the previous
properties is the following: $B\left( d\right) =
\frac{1}{d^{4}}-\frac{2}{d^{2}}.$ This function has its minimum at
$d^\star = 1$, $B\left( d^\star \right) = -1.$ For a group of three
particles $p_i=(x_i, y_i,z_i) \in \mathbb{R}^3$, $i=1,2,3$, with the
Euclidean metric $\sqrt{d_{ij}}$ $=$ $d\left(p_i,p_j\right)$. We
consider the problem of determining the orbits of a system of two given
particles and one free under potential, hereafter shall be called
$B_2$, that is, $ B_2\left( p_1, p_2 , p_3 \right) = \sum_{1 \leq i < j
\leq 3} B\left(\sqrt{d_{ij}}\right) = \sum_{1 \leq i < j \leq 3} \left(
\frac{1}{\left(d_{ij}\right)^{2}}-\frac{2}{d_{ij}} \right).$

%The well know Lennard-Jones potential (LJ 12-6) is the following
%$$ J(d)= \frac{1}{d^{12}}-\frac{2}{d^{6}},
%$$
%\noindent which has its minima at $d^\star=1$, $J\left( d^\star
%\right)=-1$. For three particles with the Euclidian metric one
%has
%\begin{eqnarray*} J_E\left( p_1, p_2 , p_3 \right) =  \sum_{1
%\leq i < j \leq 3}
%\frac{1}{\left(d_{ij}\right)^{6}}-\frac{2}{\left(d_{ij}\right)^{3}}.
%\end{eqnarray*}
%
%\noindent But using the metric
%$$\sqrt[6]{(x_i-x_j)^2+(y_i-y_j)^2+(z_i-z_j)^2}$$
%\noindent the potential LJ(12-6) yields $B$, i.e.,
%$$J_{\sqrt[6]{\cdot}}\left( p_1, p_2 , p_3 \right) = B\left( p_1,
%p_2 , p_3 \right).$$

Without lost of generality, we can assume  $p_1=(-\frac{l}{2}, 0, 0),
p_2=(\frac{l}{2}, 0, 0)$ where $l>0$ is the distance for a given system
of two fixed particles. Let be $p=(x, y, z) \in \mathbb{R}^3$ a free
particle, then the complete potential of this system is

\noindent $ B_2\left( x, y ,z \right) = \frac{1}{\left((x + \frac{l}{2}
)^2+y^2+z^2\right)^{2}}-\frac{2}{(x+\frac{l}{2})^2+y^2+z^2} +
\frac{1}{\left((x-\frac{l}{2})^2+y^2+z^2\right)^{2}}-\frac{2}{(x-\frac{l}{2})^2+y^2+z^2}
+ K_l,$

\noindent where $K_l= \frac{1}{  l^4 } - \frac{2}{l^2}$ is the
corresponding potential between $p_1$ and $p_2$, which is constant.

An orbit of $B_2$ of value $G$ is the set $\{(x, y, z) \in \mathbb{R}^3
\, | \, B_2\left( x, y ,z \right) = G\}$, $ G  \in [m_l,\infty)$ where
$l>0$, and $ m_l = \min_{(x,y,z) \in \mathbb{R}^3} B_2(x,y,z).$

The symmetry of $B_2\left( x, y ,z \right)$ with respect to second and
third axis allows the reduction of the problem to $\mathbb{R}^2$.
Moreover, $B_2\left( x, y \right) = B_2\left( x, -y \right) = B_2\left(
-x, -y \right) = B_2\left( -x, y \right).$

Therefore, it is only necessary to consider the orbits of $B_2$ of
value $G$ as the sets $O(l,G)$ = $\{(x, y) \in \mathbb{R^+}^2 \, | \,
B_2\left( x, y \right) = G\}$, $l>0$, $G  \in [m_l,\infty)$. The
corresponding 3-d model can be constructed by appropriate rotations.

Our methodology is based on:
%\begin{itemize}
    (1) The algebraic methods of Cardano and Ferrari;
    (2)  For a given $G=G_x$, the equation
     $B_2(x,y)=G$ yields a third or fourth
degree polynomial with coefficients in the ring $\mathbb{R}[x]$, that
we solve to obtain an exact root $r(x,G, l)$;
    (3) If $r(x, G, l) \geq 0 $ is a root, then $O(l,G)=
\left\{ (x,y) \, | \, x > 0, y=\sqrt{r(x, G, l)} \right\}. $
% \end{itemize}
We state now the main proposition
\begin{proposition}~\label{proposition:Orbit_of B2}
Given a system of two particles, $p_1=(-\frac{l}{2}, 0, 0),
p_2=(\frac{l}{2}, 0, 0)$. The orbits $O(l,G)$ of $B_{2}$, for $G \in
[m_l,\infty)$, correspond to the positive roots of the third and fourth
degree polynomial obtained from the equation:
\begin{equation}
B_2\left( x, y \right) - G = 0.~\label{eq:Level_set_B2}
\end{equation}

\begin{proof}
The equation~\ref{eq:Level_set_B2} gives:

\noindent $0=\left( \left( x-\frac{l}{2}\right) ^{2}+y^{2}\right) ^{2}
-
2\left( \left( x+%
\frac{l}{2}\right) ^{2}+y^{2}\right) \left( \left( x-\frac{l}{2}\right)
^{2}+y^{2}\right) ^{2}+$ $\left( \left( x+\frac{l}{2}\right)
^{2}+y^{2}\right) ^{2}-2$

\noindent $\left( \left( x- \frac{l}{2}\right) ^{2}+y^{2}\right) \left(
\left( x+\frac{l}{2}\right) ^{2}+y^{2}\right) ^{2}-$ $\left( K_l
+G\right) \left( \left( x+\frac{l}{2}\right) ^{2}+y^{2}\right)
^{2}\left( \left( x-\frac{l}{2}\right) ^{2}+y^{2}\right) ^{2}).$

It has power on $x$ or $y$ as $8,6,4,2$ when $\left( K_l+G\right) \neq
0$ and it has power on $x$ or $y$ as $6,4,2$ when $\left( K_l +G\right)
=0.$

We let now $u=y^{2}$ the following third and fourth degree equations
are obtained:
\begin{eqnarray}
0 = \left(
1024Gx^{2}+256G+1280+1024x^{2}\right) \allowbreak u^{3} &+&~\nonumber \\
 \left( 352+256Gx^{2}+3328x^{2} +
 1536Gx^{4}+1536x^{4}+96G  \right) u^{2} &+& ~\nonumber  \\
  \left(
-256Gx^{4}+1024Gx^{6}-576x^{2}-48 -  64Gx^{2}+2816x^{4}+16G+1024x^{6}
\right)
 u -15 &+&  ~\nonumber  \\
G-16Gx^{2}+96Gx^{4} - 256Gx^{6}+256Gx^{8} -
848x^{2}-672x^{4}+768x^{6}+256x^{8},~\label{eq:U_cubic_gral} \\
%\end{eqnarray}
%\begin{eqnarray}
 0 =  \left( 256G+256 K_l\right) u^{4} &+& ~\nonumber  \\
\left(
1024Gx^{2}+256G+1280+1024x^{2}\right) \allowbreak u^{3} &+&~\nonumber \\
\left( 352+256Gx^{2}+3328x^{2} +
 1536Gx^{4}+1536x^{4}+96G \right) u^{2} &+& ~\nonumber  \\
  \left(
-256Gx^{4}+1024Gx^{6}-576x^{2}-48 -  64Gx^{2}+2816x^{4}+16G+1024x^{6}
\right)
 u -15 &+&  ~\nonumber  \\
G - 16Gx^{2}+96Gx^{4} - 256Gx^{6}+256Gx^{8} -
848x^{2}-672x^{4}+768x^{6}+256x^{8}.~\label{eq:U_fourth_gral}
\end{eqnarray}

The roots of ~\ref{eq:U_cubic_gral} and ~\ref{eq:U_fourth_gral} are
constructed by the methods of Cardano and Ferrari in the complex plane
$\mathbb{C}$. Therefore, $O(l,G)$ $=$ $\{ (x,y) \, | \, x
> 0, y=\sqrt{r(x, G, l)}, r(x, G, l)$ $\geq 0 \}.$
\end{proof}
\end{proposition}

\begin{remark}
The construction of the roots is not done by mathematical software, but
it is done by finding and replacing the parameters of the formulas of
the methods of Cardano and Ferrari with the coefficients of the
polynomials of the equations~\ref{eq:U_cubic_gral}
and~\ref{eq:U_fourth_gral}. See Appendix 1.
\end{remark}

\begin{proposition}
For the system of the previous proposition, when $l=1$, we have
\begin{enumerate}
    \item $m_1=-3$.
    \item $O(1,-3) = \{ (0,\frac{\sqrt{3}}{2}) \}$. It is a
    point the root of fourth degree equation for $G_1=-3.$
    \item There are two orbits from the roots of fourth degree equation
 with  $G_1 \in \lbrack m,-1)$.
    \item There is one orbit from the roots of fourth degree equation with  $G_1\in (-1,\infty ).$
\end{enumerate}
\begin{proof}
It follows by direct substitution of $l=1$ in the resulting roots of
the previous proposition.
\end{proof}
\end{proposition}

\begin{remark}
Also for $l=1$ the figure~\ref{fig:B2_Orbits} depicts some examples of
the orbits. The free point and the other two particles form an
equilateral triangle of size $1$, and minimum is
$B_2(0,\frac{\sqrt{3}}{2}) = -3=m_1$.  There is one orbit from the
roots of third degree equation with $G_1=-1$, the details are given in
section~\ref{sc:Results}. It is easy to prove that polynomials $
Ax^{2k}+Bx^{k}+C,$ $Ax^{3k}+Bx^{2k}+Cx^{k}+D,$
$Ax^{4k}+Bx^{3k}+Cx^{2k}+Dx^{k}+E$ $k>1$ are solvable by radicals.
\end{remark}

%%%%%%%%%%%%%%%%%%%%%%%%%%%%%%%%%%%%%%%%%%%%%%%%%%%%%%%%%%%%%%%%%%%%%%%
\section{Solving Polynomial by radicals}~\label{sc:Polynomial}

The Galois theory is the algebraic framework  for the study of roots of
polynomials. It focuses on the construction of formul\ae\ using
radicals instead of in numerical estimations. For numerical estimation
there is for instance the well know method of Newton--Raphson. It is
known that there is no a general method or formula for finding the
roots of a polynomial with degree $n\geq 5$.
%Instead, the
%Galois theory is about the structure and characteristics of the groups
%of polynomials that can be solved by a formula using radicals.
Here, we are interested in applying the known methods of Cardano and
Ferrari for polynomials of degre 3 and 4 which are solvable by
radicals. Appendix 1 depicts the formulas of Cardano and Ferrari.

%That is, for a polynomial, we want to write its roots by means
%formul\ae \ involving its coefficients, arithmetic operations, and
%radicals.
%
% However, there are
%general formulas for polynomials with coefficient in $\mathbb{R}$
%when $n\leq 4$.

%The Babylonians at 1600 AC knew that the quadratic polynomial
%$f(x)=x^{2}+2px+q$ is solvable by the completing squares, by
%writing $f(x)=(x+p)^{2}+q-p^{2}$, the roots are given by
%$-p\pm\sqrt{p^{2}-q}$.

The cubic polynomial $f(x)=x^{3}+ax^{2}+bx+c$, was solved at the 16th
century by more complicated formula found simultaneously by Ferro and
Tartaglia. For the fourth degree polynomial
$f(x)=x^{4}+ax^{3}+bx^{2}+cx+d$, Ferrari provided a procedure. These
methods were published by Cardano in the {\it Ars Magna} at 1545. In
18th century, Lagrange unified these methods for a polynomial with
degree $n\leq 4$ using what now is know as the resolvent of Lagrange.
This method involves an auxiliary equation with a polynomial with
degree less than one, i.e., it uses, for example, a third degree
polynomial to solve the fourth degree polynomial. In fact, this is the
procedure for solving the quadratic polynomial. However, this method
fails for the polynomials of degree $5$, because the auxiliary
polynomial is of degree $6$. Ruffini at 1799, and Abel at 1824 proved
that there is not a general formula using radicals for finding the
roots of quintic polynomial. Galois by 1832 showed how to associate to
each polynomial $f(x)$ a subgroup $Gal(f)$ of the symmetric group, call
the Galois group of $f(x)$, and established the following result, for
details see for instance~\cite{bk:Rotman}
\begin{theorem} (Galois) A polynomial $f$ is soluble by radicals if and only if its group
$Gal(f)$ is soluble.
\end{theorem}
As an application of these algebraic techniques of Cardano and Ferrari,
we have the following result.

\begin{proposition}~\label{proposition:Optimal_points_B2}
Given a system of two particles, $p_1=(-\frac{l}{2}, 0, 0),
p_2=(\frac{l}{2}, 0, 0)$, the critical points of $B_2\left( x, y
\right)$ are the following
\begin{enumerate}
    \item  $(0,0)$ $\forall l \geq 0$.
    \item  $(\pm x^\star,0)$, $x^\star$ is a positive root of the polynomial obtained from substituting $y=0$ in
    $\frac{\partial}{\partial x}B_2\left( x, y \right)$ with $l >
    2$. The points $(\pm x^\star,0)$ are collinear with $p_1, p_2.$
    \item  $(0, \pm y^\star)$ where $y^\star$ is a
    positive root of the polynomial obtained from substituting $x=0$ in
    $\frac{\partial}{\partial y}B_2\left( x, y \right)$.
    The points $(0, \pm y^\star)$ correspond to the oppositive vertex of an
    isosceles triangle of the side with vertices $p_1, p_2.$
\end{enumerate}
\begin{proof}
The polynomial to solve comes from the first optimality condition:
$\frac{\partial}{\partial x}B_2\left( x, y \right)$ $=0$ and
$\frac{\partial}{\partial y}B_2\left( x, y \right)=0,$ where

\begin{eqnarray*}
  \frac{\partial }{\partial x}B_{2}\left( x,y\right) &=& -4\frac{(x \pm \frac{l}{2}) }{\left( (x \pm
\frac{l}{2})^{2}+y^{2}\right) ^{3}} + \frac{4(x \pm \frac{l}{2})}{
\left( (x \pm \frac{l}{2})^{2}+y^{2}\right) ^{2}},
 \\
  \frac{\partial }{\partial y}B_{2}\left( x,y\right) &=& -4\frac{y}{\left( (x \pm \frac{l}{2})^{2}+y^{2}\right)
^{3}}+\frac{4y}{\left( (x \pm \frac{l}{2} )^{2}+y^{2}\right) ^{2}}.
\end{eqnarray*}
\noindent Note that from the previous equations, we have that $\nabla
B_2(0,0)=0,$ for all $l>0.$

\noindent Using that $y=0$ gives $\frac{\partial}{\partial
y}B_2(x,0)=0$, the equation for $\frac{\partial}{\partial
x}B_2(u^\frac{1}{2},0)=0$ ($x=u^2$ is changed) is
\begin{eqnarray}
-64u^{3}+\left( 64-16l^{2}\right) u^{2} +
 \left(
20l^{4}+160l^{2}\right) u+\allowbreak \left( 20l^{4}-3l^{6}\right) &=&
0.~\label{eq:poly_U}
\end{eqnarray}

The positive roots of the previous third degree equation give the
optimal points. In a similar way, using that $x=0$ gives
$\frac{\partial}{\partial x}B_2(0, u^\frac{1}{2})=0$, the equation for
$\frac{\partial}{\partial x}B_2(0, u^\frac{1}{2}) $ $=0$ gives
\begin{equation}
4y^2+l^2-4=0 ~\label{eq:poly_Y}
\end{equation}

The real roots of the previous second degree equation give the optimal
points for $l \in [0, 2)$. The interval of $l$ come from the
restriction of the discriminant of the quadratic equation.
\end{proof}
\end{proposition}

\section{Levels set of the potential $B_2$}~\label{sc:Results}
%%%%%%%%%%%%%%%%%%%%%%%%%%%%%%%%%%%%%%%%%%%%%%%%%%%%%%%%%%%%%%%%%%%%
%\subsection{Numerical Results}~\label{sc:Numerical_Results}
%
%Here the configuration of the fixed particles correspond to $l=1$.
%For $G=-1. 9305$ the figure~\ref{fig:Y1_14CurvasNivel_V2} depicts
%a numerical approximation for the orbit $O(1,-1. 9305).$
%Figure~\ref{fig:Y1_14Potential} illustrates the corresponding
%values of the potential. The expected result is not the constant
%line at $G=-1. 9305$.
%
%\begin{figure}[tbp]
%\centering \psfig{figure=Pot2_CnivelY1_14.ps,height=1.2in} \caption{The
%dots depict the numerical approximation of the orbit $O(1,-1. 9305)$.}
%~\label{fig:Y1_14CurvasNivel_V2}
%\end{figure}
%
%
%
%\begin{figure}[tbp]
%\centerline{ \psfig{figure=Pot2_CnivelY1_14_B2.ps,width=2.0in,height=1.0in}}%
%\caption{The Potential of the numerical approximation of the orbit
%$O(1,-1. 9305).$} ~\label{fig:Y1_14Potential}
%\end{figure}
% \subsection{Results}

The proposition~\ref{proposition:Optimal_points_B2} gives similar
results for $B_2$ instead of LJ 12-6 to the triangular and collinear
choreographes of the 3-body system discussed
in~\cite{phyAstr:Corbera2004}. We remark that our approach is done by
basic analysis and algebraic techniques for the static case.

%The next proposition provides the asymptotics for the isosceles
%triangular configurations.
\begin{proposition}~\label{proposition:Triangle_optimal_points_B2}
Given a system of two particles of the
proposition~\ref{proposition:Optimal_points_B2}, and $l \in (2,0)$. If
a free particle is at $(0,y)$, where $y = \pm \frac{1}{2}\sqrt{2^2 -
l^2 }$. Then the free particle is on a local point. Moreover, from the
free particle's point of view, the two particles act as one virtual
particle of double $B$ potential when $l \to 0$.

\begin{proof}
Without loss of generality, let $y_1 = \frac{1}{2}\sqrt{2^2 - l^2 }$.
The free particle is perpendicular to middle point of the side with
vertices $(p_1, p_2).$  The result follows immediately from
proposition~\ref{proposition:Optimal_points_B2} and the
equation~\ref{eq:poly_Y}, which corresponds to the condition $\nabla
B_2(0,y_1)=0$. Also, the free particle position goes to $(0,2)$ when $l
\to 0$, which is the double of the optimal distance ($d^\star=1$) of
the function $B$.
\end{proof}
\end{proposition}

\begin{proposition}~\label{proposition:Triangle_equilataral_points_B2}
Given a system of two particles of the
proposition~\ref{proposition:Optimal_points_B2}, with $l=1$.  Then the
minimal points of a free particle is at $(0,y)$, where $y = \pm
\frac{\sqrt{3}}{2}$. The particles are the vertices of an equilateral
triangle.

\begin{proof}
Without loss of generality, let  $y_1 = \frac{1}{2}\sqrt{3}$, then the
distance between the three particles is 1. The Hessian of the system at
$(0,y_1)$ is
$$ |\nabla^2
B_2(0,\frac{\sqrt{3}}{2})|=-64\left(-3\right).$$ The optimality follows
from proposition~\ref{proposition:Optimal_points_B2} and $|\nabla^2
B_2(0,y_1)|>0$.
\end{proof}
\end{proposition}

\begin{proposition}~\label{optimal_orbit_point_B2}
Given a system of two particles of the
proposition~\ref{proposition:Optimal_points_B2}, and $l \in (2,0)$.
Then the orbit of $B_2$ of value $G_y=B_2(0,y)$ is the pointed set
$O(l,G_y)$= $\{(0, y) \in \mathbb{R^+}^2 \}$ where $y =
\frac{1}{2}\sqrt{2^2 - l^2 }.$

\begin{proof}
By the symmetry the space is restricted to $\mathbb{R^+}^2$. The
Hessian of the system at $(0,y)$ is
$$ |\nabla^2
B_2(0,\frac{1}{2}\sqrt{2^2 - l^2 })|=-64l^{2}\left( l^{2}-4\right).$$

The polynomial  $-64l^{2}\left( l^{2}-4\right)$ is strictly positive
for $l \in (0,2)$. This means that the positive root of the
equation~\ref{eq:poly_Y} corresponds to the optimality conditions
$\nabla B_2(0,y)=0$ and $|\nabla^2 B_2(0,y)|>0$. Therefore $O(l,G_y)$
is the minimal point $(0,y)$.
\end{proof}
\end{proposition}

\begin{figure}[tbp]
\begin{minipage}{0.445 \textwidth}
\centerline{ \psfig{figure=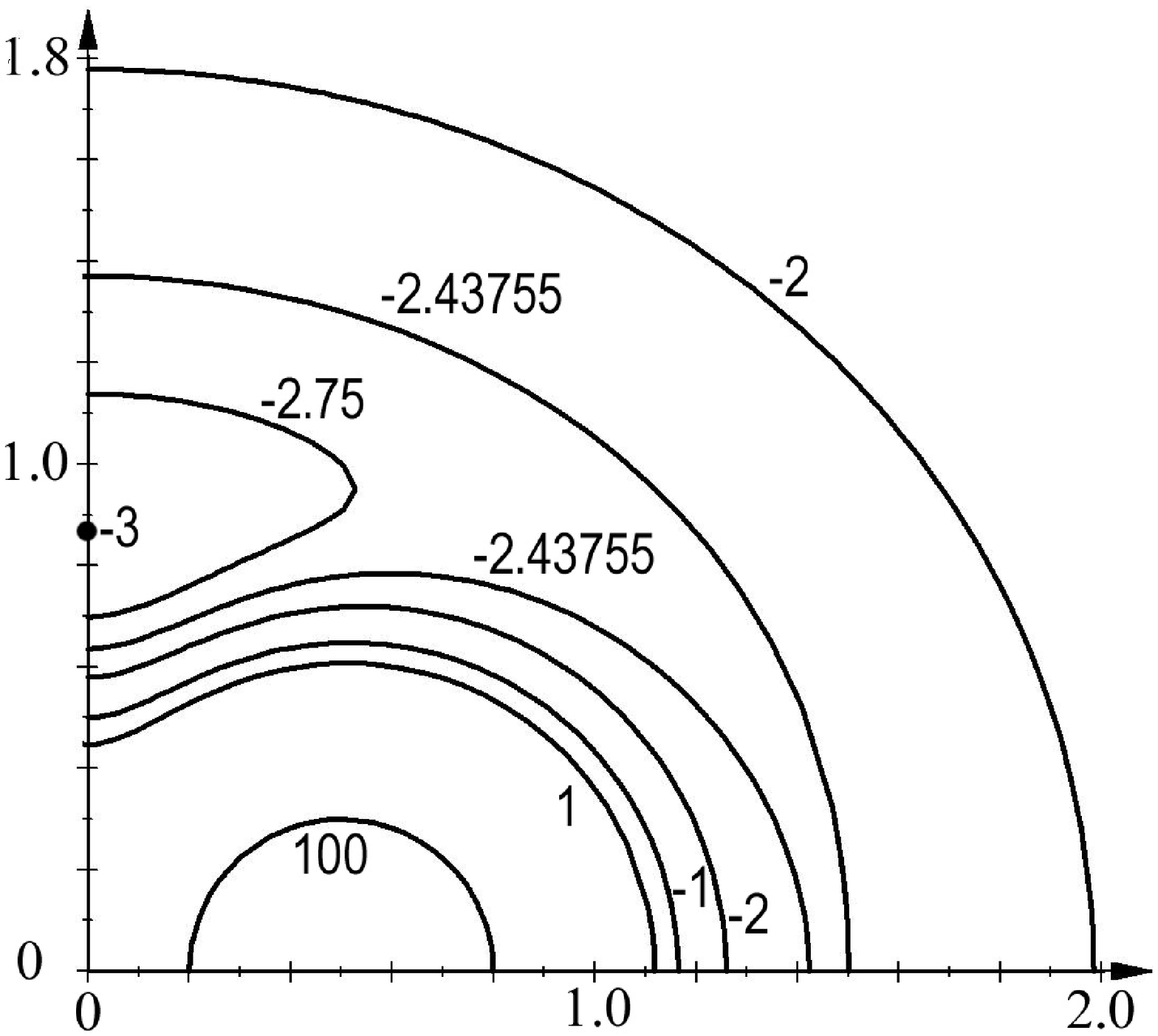,width=2.0in,height=1.75in}
} \caption{The true orbits for $O(1,-3)$, $O(1,-2.75)$,
$O(1,-2.43755)$, $O(1,-2)$ , $O(1,-1)$ , $O(1,0)$ , and
$O(1,100).$}~\label{fig:B2_Orbits}
\end{minipage}
\hspace{0.05 \textwidth}
\begin{minipage}{0.45 \textwidth}
\centerline{\psfig{figure=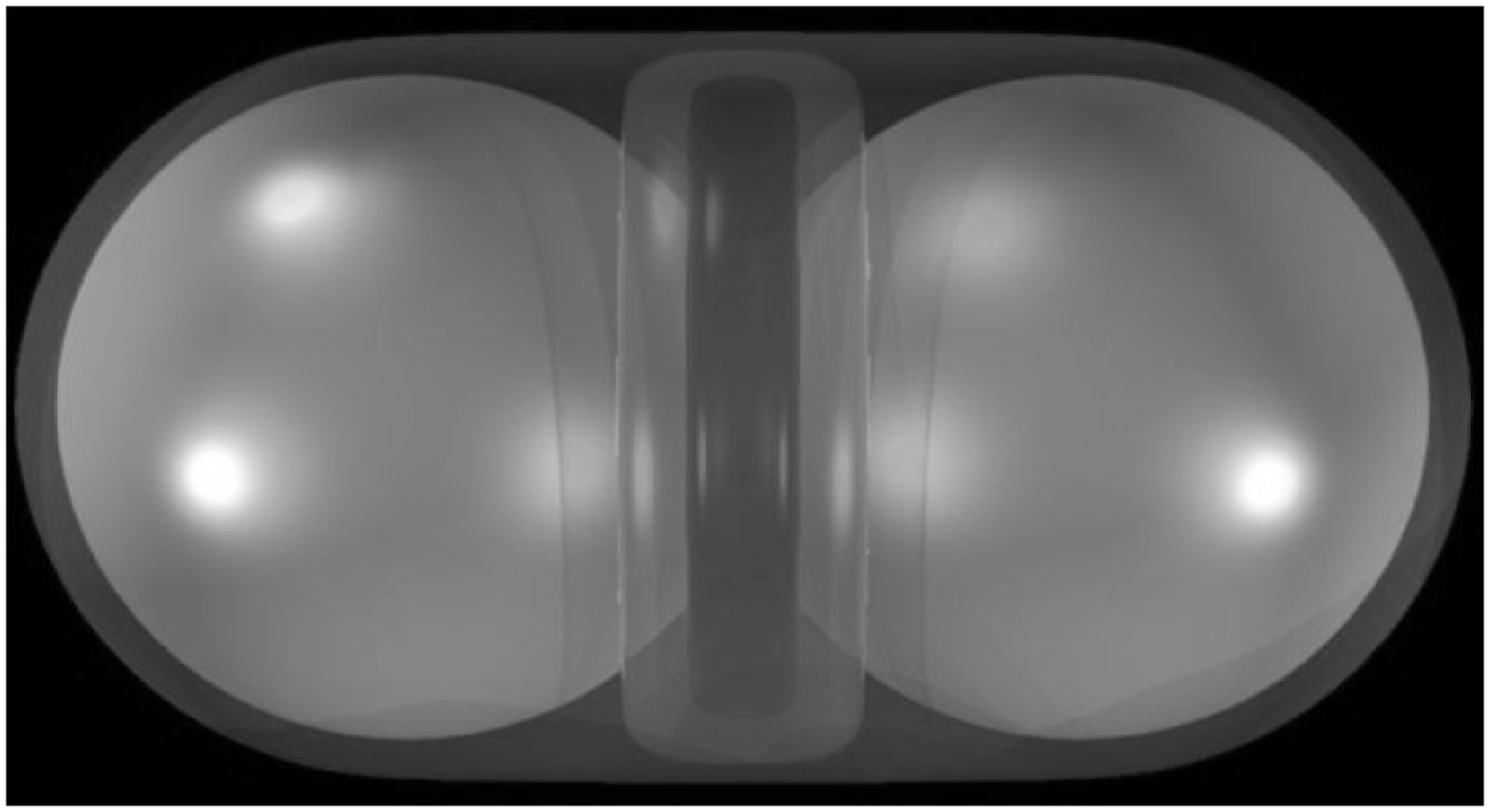,width=2.1in,height=1.45in}}
\caption{3-d model of the level sets of $B_2(x,y,z)=G,$ where $G
\in[3,\infty)$.} ~\label{fig:B2_3DModel}
\end{minipage}
\end{figure}

For $G=-3$ the four roots of fourth degree
equation~\ref{eq:U_fourth_gral} (with $l=1$) are

$r_{1}\left( x\right)
=\frac{1}{4}-x^{2}-\frac{1}{2}\sqrt{4x^{2}-4ix+1}$, $r_{2}\left(
x\right) =\frac{1}{2}\sqrt{4x^{2}-4ix+1}-x^{2}+\frac{1}{4}$

$r_{3}\left( x\right)
=\frac{1}{4}-x^{2}-\frac{1}{2}\sqrt{4x^{2}+4ix+1}$, $r_{4}\left(
x\right) =\frac{1}{2}\sqrt{4x^{2}+4ix+1}-x^{2}+\frac{1}{4}.$

Only for $x=0,$  $r_{2}\left( 0 \right)=\frac{3}{4} > 0$, therefore the
orbit of $O(1,-3)$ is the point set $\{(0,\sqrt{\frac{3}{4}})\}$. Which
is the optimal point of the equilateral triangle of size $=1$. Here,
our method gives the result without appealing to the optimal conditions
as in proposition~\ref{proposition:Triangle_equilataral_points_B2} and
in proposition~\ref{optimal_orbit_point_B2}.

The detailed analysis of the third and fourth degree equations of
proposition~\ref{proposition:Orbit_of B2} for $l=1$ was done for
verifying the following:
\begin{enumerate}
    \item $m_1=-3$.
    \item $O(1,-3) = \{ (0,\frac{\sqrt{3}}{2}) \}$, i.e., it is a
    point the root of fourth degree equation for $G_1=-3.$
    \item There are two orbits from the roots of fourth degree equation
    with  $G_1 \in \lbrack m,-1)$
    \item There is one orbit from the roots of third degree equation with  $G_1=-1$.
    \item There is one orbit from the roots of fourth degree equation with  $G_1\in (-1,\infty ).$
\end{enumerate}

Figure~\ref{fig:B2_Orbits} depicts some examples of the orbits that
comply the previous statements. Figure~\ref{fig:B2_3DModel} depicts a
3-d model of the level sets for $l=1$ and $G\in[-3,\infty)$.
%%%%%%%%%%%%%%%%%%%%%%%%%%%%%%%%%%%%%%%%%%%%%%%%%%%%%%%%%%%%%%%%%%%%%%%%%%%%%%%%%
%%%%%%%%%%%%%%%%%%%%%%%%%%%%%%%%%%%%%%%%%%%%%%%%%%%%%%%%%%%%%%%%%%%%%%%%%%%%%%%%%
\section{Conclusions}~\label{sc:conclusions and future work}

To our knowledge this is the first totally complete description of the
level sets of a system of two particles under a Van der Waals
Potential.
%The application of the research on solvable polynomial
%for a Galois Groups is promising~\cite{MC:Ward1999} to apply as
%here.

We have presented a novel approach for the study of orbits of systems
of non interactive particles. We tackle the problem by reducing the
analysis of the equipotential zones to the one of finding the roots of
certain polynomials.
% The structure of such
%polynomial allows the use of algebraic methods based on the
%solvability of the associated Galois group.
A constructive method provided by the methods of Cardano and Ferrari,
yields a  complete factorization of the polynomials and consequently an
analytical description of equipotential  zones.

\noindent \textbf{Acknowledgement.} Thanks to the Department of Basic
Sciences at UAM, Azcapotzalco for supporting this research.

\bigskip

\noindent \textbf{\large Appendix 1. Formulae of Cardano and Ferrari
for~\ref{eq:Level_set_B2}}
%}~\label{CardanFerrariFormulae}

\noindent The third degree polynomial is obtained from the
equation~\ref{eq:Level_set_B2}. This equation gives the third degree
equation~\ref{eq:U_cubic_gral} with $u=y^{2}$. Its polynomial
corresponds to $Ax^{3}+Bx^{2}+Cx+D=0.$ And its root come from the
following formula of Cardano:

$x_{1}=((-\frac{1}{2}\allowbreak (\frac{2}{27}((B)\allowbreak
(A)^{-1})^{3}-\frac{1}{3}((B)\allowbreak (A)^{-1})\allowbreak
((C)\allowbreak (A)^{-1})+\allowbreak ((D)\allowbreak
(A)^{-1}))+\allowbreak (\frac{1}{27}((C)\allowbreak
(A)^{-1}-\frac{1}{3}\allowbreak ((B)\allowbreak
(A)^{-1})^{2})^{3}+\allowbreak \frac{1}{4}(\frac{2}{27}((B)\allowbreak
(A)^{-1})^{3}-\frac{1}{3}\allowbreak ((B)\allowbreak
(A)^{-1})\allowbreak ((C)\allowbreak (A)^{-1})+\allowbreak ((D)$

$\allowbreak (A)^{-1}))^{2})^{\frac{
1}{2}})^{\frac{1}{3}})\allowbreak +((-\frac{1}{2}\allowbreak (\frac{2}{27}%
((B)\allowbreak (A)^{-1})^{3}-\frac{1}{3}((B)\allowbreak
(A)^{-1})\allowbreak ((C)\allowbreak (A)^{-1})+\allowbreak
((D)\allowbreak
(A)^{-1}))-\allowbreak (\frac{1}{27}((C)\allowbreak (A)^{-1}-\frac{1}{3}%
\allowbreak ((B)\allowbreak (A)^{-1})^{2})^{3}+\allowbreak \frac{1}{4}(\frac{%
2}{27}((B)\allowbreak (A)^{-1})^{3}-\frac{1}{3}((B)\allowbreak
(A)^{-1})((C)\allowbreak (A)^{-1})+((D)\allowbreak (A)^{-1}))^{2})^{\frac{1}{%
2}})^{\frac{1}{3}})-\allowbreak \frac{1}{3}((B)\allowbreak (A)^{-1})$

The fourth degree polynomial is obtained from the
equation~\ref{eq:Level_set_B2}. For the equation
\ref{eq:U_fourth_gral}, with polynomial type
$Ax^{4}+Bx^{3}+Cx^{2}+Dx+E=0$ the four corresponding roots come from
the following four formulae of Ferrari.

The first root is

$x_{1}=-\frac{1}{4}(B)\allowbreak (A)^{-1}+\allowbreak \frac{1}{2}%
(\allowbreak (((-\frac{3}{8}(B)^{2}\allowbreak (A)^{-2}+\allowbreak
(C)\allowbreak (A)^{-1})+\allowbreak 2(-\frac{5}{6}\allowbreak (-\frac{3}{8}%
(B)^{2}\allowbreak (A)^{-2}+\allowbreak (C)\allowbreak
(A)^{-1})+\allowbreak (((-\frac{1}{2}(-\frac{1}{108}\allowbreak
(-\frac{3}{8}(B)^{2}\allowbreak
(A)^{-2}+\allowbreak (C)\allowbreak (A)^{-1})^{3}+\allowbreak \frac{1}{3}(-%
\frac{3}{8}(B)^{2}\allowbreak (A)^{-2}+\allowbreak (C)\allowbreak
(A)^{-1})\allowbreak (-\frac{3}{256}(B)^{4}\allowbreak
(A)^{-4}+\allowbreak
\frac{1}{16}(C)\allowbreak (B)^{2}\allowbreak (A)^{-3}-\allowbreak \frac{1}{4%
}(B)\allowbreak (D)\allowbreak (A)^{-2}+\allowbreak (E)\allowbreak
(A)^{-1})-\allowbreak \frac{1}{8}(\frac{1}{8}(B)^{3}\allowbreak
(A)^{-3}-\allowbreak \frac{1}{2}(B)\allowbreak (C)\allowbreak
(A)^{-2}+\allowbreak (D)\allowbreak (A)^{-1})^{2})\allowbreak +(\frac{1}{4}%
\allowbreak (-\frac{1}{108}\allowbreak (-\frac{3}{8}(B)^{2}\allowbreak
(A)^{-2}+\allowbreak (C)\allowbreak (A)^{-1})^{3}+\allowbreak \frac{1}{3}(-%
\frac{3}{8}(B)^{2}\allowbreak (A)^{-2}+\allowbreak (C)\allowbreak
(A)^{-1})\allowbreak (-\frac{3}{256}(B)^{4}\allowbreak
(A)^{-4}+\allowbreak
\frac{1}{16}(C)\allowbreak (B)^{2}\allowbreak (A)^{-3}-\allowbreak \frac{1}{4%
}(B)\allowbreak (D)\allowbreak (A)^{-2}+\allowbreak (E)\allowbreak
(A)^{-1})-\allowbreak \frac{1}{8}(\frac{1}{8}(B)^{3}\allowbreak
(A)^{-3}-\allowbreak \frac{1}{2}(B)\allowbreak (C)\allowbreak
(A)^{-2}+\allowbreak (D)\allowbreak (A)^{-1})^{2})^{2}+\allowbreak \frac{1}{%
27}\allowbreak (-\frac{1}{12}\allowbreak
(-\frac{3}{8}(B)^{2}\allowbreak
(A)^{-2}+\allowbreak (C)\allowbreak (A)^{-1})^{2}-\allowbreak (-\frac{3}{256}%
(B)^{4}\allowbreak (A)^{-4}+\allowbreak \frac{1}{16}(C)\allowbreak
(B)^{2}\allowbreak (A)^{-3}-\allowbreak \frac{1}{4}(B)\allowbreak
(D)\allowbreak (A)^{-2}+\allowbreak (E)\allowbreak (A)^{-1}))^{3})^{\frac{1}{%
2}}))^{\frac{1}{3}})-\allowbreak \frac{1}{3}(-\frac{1}{12}\allowbreak (-%
\frac{3}{8}(B)^{2}\allowbreak (A)^{-2}+\allowbreak (C)\allowbreak
(A)^{-1})^{2}-\allowbreak (-\frac{3}{256}(B)^{4}\allowbreak
(A)^{-4}+\allowbreak \frac{1}{16}(C)\allowbreak (B)^{2}\allowbreak
(A)^{-3}-\allowbreak \frac{1}{4}(B)\allowbreak (D)\allowbreak
(A)^{-2}+\allowbreak (E)\allowbreak (A)^{-1}))\allowbreak (((-\frac{1}{2}(-%
\frac{1}{108}\allowbreak (-\frac{3}{8}(B)^{2}\allowbreak
(A)^{-2}+\allowbreak (C)\allowbreak (A)^{-1})^{3}+\allowbreak \frac{1}{3}(-%
\frac{3}{8}(B)^{2}\allowbreak (A)^{-2}+\allowbreak (C)\allowbreak
(A)^{-1})\allowbreak (-\frac{3}{256}(B)^{4}\allowbreak
(A)^{-4}+\allowbreak
\frac{1}{16}(C)\allowbreak (B)^{2}\allowbreak (A)^{-3}-\allowbreak \frac{1}{4%
}(B)\allowbreak (D)\allowbreak (A)^{-2}+\allowbreak (E)\allowbreak
(A)^{-1})-\allowbreak \frac{1}{8}(\frac{1}{8}(B)^{3}\allowbreak
(A)^{-3}-\allowbreak \frac{1}{2}(B)\allowbreak (C)\allowbreak
(A)^{-2}+\allowbreak (D)\allowbreak (A)^{-1})^{2})\allowbreak +(\frac{1}{4}%
\allowbreak (-\frac{1}{108}\allowbreak (-\frac{3}{8}(B)^{2}\allowbreak
(A)^{-2}+\allowbreak (C)\allowbreak (A)^{-1})^{3}+\allowbreak \frac{1}{3}(-%
\frac{3}{8}(B)^{2}\allowbreak (A)^{-2}+\allowbreak (C)\allowbreak
(A)^{-1})\allowbreak (-\frac{3}{256}(B)^{4}\allowbreak
(A)^{-4}+\allowbreak
\frac{1}{16}(C)\allowbreak (B)^{2}\allowbreak (A)^{-3}-\allowbreak \frac{1}{4%
}(B)\allowbreak (D)\allowbreak (A)^{-2}+\allowbreak (E)\allowbreak
(A)^{-1})-\allowbreak \frac{1}{8}(\frac{1}{8}(B)^{3}\allowbreak
(A)^{-3}-\allowbreak \frac{1}{2}(B)\allowbreak (C)\allowbreak
(A)^{-2}+\allowbreak (D)\allowbreak (A)^{-1})^{2})^{2}+\allowbreak \frac{1}{%
27}\allowbreak (-\frac{1}{12}\allowbreak
(-\frac{3}{8}(B)^{2}\allowbreak
(A)^{-2}+\allowbreak (C)\allowbreak (A)^{-1})^{2}-\allowbreak (-\frac{3}{256}%
(B)^{4}\allowbreak (A)^{-4}+\allowbreak \frac{1}{16}(C)\allowbreak
(B)^{2}\allowbreak (A)^{-3}-\allowbreak \frac{1}{4}(B)\allowbreak
(D)\allowbreak (A)^{-2}+\allowbreak (E)\allowbreak (A)^{-1}))^{3})^{\frac{1}{%
2}}))^{\frac{1}{3}})^{-1}))^{\frac{1}{2}})-\allowbreak (-(3\allowbreak (-%
\frac{3}{8}(B)^{2}\allowbreak (A)^{-2}+\allowbreak (C)\allowbreak
(A)^{-1})+\allowbreak 2\allowbreak (-\frac{5}{6}\allowbreak (-\frac{3}{8}%
(B)^{2}\allowbreak (A)^{-2}+\allowbreak (C)\allowbreak
(A)^{-1})+\allowbreak (((-\frac{1}{2}(-\frac{1}{108}\allowbreak
(-\frac{3}{8}(B)^{2}\allowbreak
(A)^{-2}+\allowbreak (C)\allowbreak (A)^{-1})^{3}+\allowbreak \frac{1}{3}(-%
\frac{3}{8}(B)^{2}\allowbreak (A)^{-2}+\allowbreak (C)\allowbreak
(A)^{-1})\allowbreak (-\frac{3}{256}(B)^{4}\allowbreak
(A)^{-4}+\allowbreak
\frac{1}{16}(C)\allowbreak (B)^{2}\allowbreak (A)^{-3}-\allowbreak \frac{1}{4%
}(B)\allowbreak (D)\allowbreak (A)^{-2}+\allowbreak (E)\allowbreak
(A)^{-1})-\allowbreak \frac{1}{8}(\frac{1}{8}(B)^{3}\allowbreak
(A)^{-3}-\allowbreak \frac{1}{2}(B)\allowbreak (C)\allowbreak
(A)^{-2}+\allowbreak (D)\allowbreak (A)^{-1})^{2})\allowbreak +(\frac{1}{4}%
\allowbreak (-\frac{1}{108}\allowbreak (-\frac{3}{8}(B)^{2}\allowbreak
(A)^{-2}+\allowbreak (C)\allowbreak (A)^{-1})^{3}+\allowbreak \frac{1}{3}(-%
\frac{3}{8}(B)^{2}\allowbreak (A)^{-2}+\allowbreak (C)\allowbreak
(A)^{-1})\allowbreak (-\frac{3}{256}(B)^{4}\allowbreak
(A)^{-4}+\allowbreak
\frac{1}{16}(C)\allowbreak (B)^{2}\allowbreak (A)^{-3}-\allowbreak \frac{1}{4%
}(B)\allowbreak (D)\allowbreak (A)^{-2}+\allowbreak (E)\allowbreak
(A)^{-1})-\allowbreak \frac{1}{8}(\frac{1}{8}(B)^{3}\allowbreak
(A)^{-3}-\allowbreak \frac{1}{2}(B)\allowbreak (C)\allowbreak
(A)^{-2}+\allowbreak (D)\allowbreak (A)^{-1})^{2})^{2}+\allowbreak \frac{1}{%
27}\allowbreak (-\frac{1}{12}\allowbreak
(-\frac{3}{8}(B)^{2}\allowbreak
(A)^{-2}+\allowbreak (C)\allowbreak (A)^{-1})^{2}-\allowbreak (-\frac{3}{256}%
(B)^{4}\allowbreak (A)^{-4}+\allowbreak \frac{1}{16}(C)\allowbreak
(B)^{2}\allowbreak (A)^{-3}-\allowbreak \frac{1}{4}(B)\allowbreak
(D)\allowbreak (A)^{-2}+\allowbreak (E)\allowbreak (A)^{-1}))^{3})^{\frac{1}{%
2}}))^{\frac{1}{3}})-\allowbreak \frac{1}{3}(-\frac{1}{12}\allowbreak (-%
\frac{3}{8}(B)^{2}\allowbreak (A)^{-2}+\allowbreak (C)\allowbreak
(A)^{-1})^{2}-\allowbreak (-\frac{3}{256}(B)^{4}\allowbreak
(A)^{-4}+\allowbreak \frac{1}{16}(C)\allowbreak (B)^{2}\allowbreak
(A)^{-3}-\allowbreak \frac{1}{4}(B)\allowbreak (D)\allowbreak
(A)^{-2}+\allowbreak (E)\allowbreak (A)^{-1}))\allowbreak (((-\frac{1}{2}(-%
\frac{1}{108}\allowbreak (-\frac{3}{8}(B)^{2}\allowbreak
(A)^{-2}+\allowbreak (C)\allowbreak (A)^{-1})^{3}+\allowbreak \frac{1}{3}(-%
\frac{3}{8}(B)^{2}\allowbreak (A)^{-2}+\allowbreak (C)\allowbreak
(A)^{-1})\allowbreak (-\frac{3}{256}(B)^{4}\allowbreak
(A)^{-4}+\allowbreak
\frac{1}{16}(C)\allowbreak (B)^{2}\allowbreak (A)^{-3}-\allowbreak \frac{1}{4%
}(B)\allowbreak (D)\allowbreak (A)^{-2}+\allowbreak (E)\allowbreak
(A)^{-1})-\allowbreak \frac{1}{8}(\frac{1}{8}(B)^{3}\allowbreak
(A)^{-3}-\allowbreak \frac{1}{2}(B)\allowbreak (C)\allowbreak
(A)^{-2}+\allowbreak (D)\allowbreak (A)^{-1})^{2})\allowbreak +(\frac{1}{4}%
\allowbreak (-\frac{1}{108}\allowbreak (-\frac{3}{8}(B)^{2}\allowbreak
(A)^{-2}+\allowbreak (C)\allowbreak (A)^{-1})^{3}+\allowbreak \frac{1}{3}(-%
\frac{3}{8}(B)^{2}\allowbreak (A)^{-2}+\allowbreak (C)\allowbreak
(A)^{-1})\allowbreak (-\frac{3}{256}(B)^{4}\allowbreak
(A)^{-4}+\allowbreak
\frac{1}{16}(C)\allowbreak (B)^{2}\allowbreak (A)^{-3}-\allowbreak \frac{1}{4%
}(B)\allowbreak (D)\allowbreak (A)^{-2}+\allowbreak (E)\allowbreak
(A)^{-1})-\allowbreak \frac{1}{8}(\frac{1}{8}(B)^{3}\allowbreak
(A)^{-3}-\allowbreak \frac{1}{2}(B)\allowbreak (C)\allowbreak
(A)^{-2}+\allowbreak (D)\allowbreak (A)^{-1})^{2})^{2}+\allowbreak \frac{1}{%
27}\allowbreak (-\frac{1}{12}\allowbreak
(-\frac{3}{8}(B)^{2}\allowbreak
(A)^{-2}+\allowbreak (C)\allowbreak (A)^{-1})^{2}-\allowbreak (-\frac{3}{256}%
(B)^{4}\allowbreak (A)^{-4}+\allowbreak \frac{1}{16}(C)\allowbreak
(B)^{2}\allowbreak (A)^{-3}-\allowbreak \frac{1}{4}(B)\allowbreak
(D)\allowbreak (A)^{-2}+\allowbreak (E)\allowbreak (A)^{-1}))^{3})^{\frac{1}{%
2}}))^{\frac{1}{3}})^{-1})+\allowbreak 2\allowbreak (\frac{1}{8}%
(B)^{3}\allowbreak (A)^{-3}-\allowbreak \frac{1}{2}(B)\allowbreak
(C)\allowbreak (A)^{-2}+\allowbreak (D)\allowbreak (A)^{-1})\allowbreak (((-%
\frac{3}{8}(B)^{2}\allowbreak (A)^{-2}+\allowbreak (C)\allowbreak
(A)^{-1})+\allowbreak 2(-\frac{5}{6}\allowbreak (-\frac{3}{8}%
(B)^{2}\allowbreak (A)^{-2}+\allowbreak (C)\allowbreak
(A)^{-1})+\allowbreak (((-\frac{1}{2}(-\frac{1}{108}\allowbreak
(-\frac{3}{8}(B)^{2}\allowbreak
(A)^{-2}+\allowbreak (C)\allowbreak (A)^{-1})^{3}+\allowbreak \frac{1}{3}(-%
\frac{3}{8}(B)^{2}\allowbreak (A)^{-2}+\allowbreak (C)\allowbreak
(A)^{-1})\allowbreak (-\frac{3}{256}(B)^{4}\allowbreak
(A)^{-4}+\allowbreak
\frac{1}{16}(C)\allowbreak (B)^{2}\allowbreak (A)^{-3}-\allowbreak \frac{1}{4%
}(B)\allowbreak (D)\allowbreak (A)^{-2}+\allowbreak (E)\allowbreak
(A)^{-1})-\allowbreak \frac{1}{8}(\frac{1}{8}(B)^{3}\allowbreak
(A)^{-3}-\allowbreak \frac{1}{2}(B)\allowbreak (C)\allowbreak
(A)^{-2}+\allowbreak (D)\allowbreak (A)^{-1})^{2})\allowbreak +(\frac{1}{4}%
\allowbreak (-\frac{1}{108}\allowbreak (-\frac{3}{8}(B)^{2}\allowbreak
(A)^{-2}+\allowbreak (C)\allowbreak (A)^{-1})^{3}+\allowbreak \frac{1}{3}(-%
\frac{3}{8}(B)^{2}\allowbreak (A)^{-2}+\allowbreak (C)\allowbreak
(A)^{-1})\allowbreak (-\frac{3}{256}(B)^{4}\allowbreak
(A)^{-4}+\allowbreak
\frac{1}{16}(C)\allowbreak (B)^{2}\allowbreak (A)^{-3}-\allowbreak \frac{1}{4%
}(B)\allowbreak (D)\allowbreak (A)^{-2}+\allowbreak (E)\allowbreak
(A)^{-1})-\allowbreak \frac{1}{8}(\frac{1}{8}(B)^{3}\allowbreak
(A)^{-3}-\allowbreak \frac{1}{2}(B)\allowbreak (C)\allowbreak
(A)^{-2}+\allowbreak (D)\allowbreak (A)^{-1})^{2})^{2}+\allowbreak \frac{1}{%
27}\allowbreak (-\frac{1}{12}\allowbreak
(-\frac{3}{8}(B)^{2}\allowbreak
(A)^{-2}+\allowbreak (C)\allowbreak (A)^{-1})^{2}-\allowbreak (-\frac{3}{256}%
(B)^{4}\allowbreak (A)^{-4}+\allowbreak \frac{1}{16}(C)\allowbreak
(B)^{2}\allowbreak (A)^{-3}-\allowbreak \frac{1}{4}(B)\allowbreak
(D)\allowbreak (A)^{-2}+\allowbreak (E)\allowbreak (A)^{-1}))^{3})^{\frac{1}{%
2}}))^{\frac{1}{3}})-\allowbreak \frac{1}{3}(-\frac{1}{12}\allowbreak (-%
\frac{3}{8}(B)^{2}\allowbreak (A)^{-2}+\allowbreak (C)\allowbreak
(A)^{-1})^{2}-\allowbreak (-\frac{3}{256}(B)^{4}\allowbreak
(A)^{-4}+\allowbreak \frac{1}{16}(C)\allowbreak (B)^{2}\allowbreak
(A)^{-3}-\allowbreak \frac{1}{4}(B)\allowbreak (D)\allowbreak
(A)^{-2}+\allowbreak (E)\allowbreak (A)^{-1}))\allowbreak (((-\frac{1}{2}(-%
\frac{1}{108}\allowbreak (-\frac{3}{8}(B)^{2}\allowbreak
(A)^{-2}+\allowbreak (C)\allowbreak (A)^{-1})^{3}+\allowbreak \frac{1}{3}(-%
\frac{3}{8}(B)^{2}\allowbreak (A)^{-2}+\allowbreak (C)\allowbreak
(A)^{-1})\allowbreak (-\frac{3}{256}(B)^{4}\allowbreak
(A)^{-4}+\allowbreak
\frac{1}{16}(C)\allowbreak (B)^{2}\allowbreak (A)^{-3}-\allowbreak \frac{1}{4%
}(B)\allowbreak (D)\allowbreak (A)^{-2}+\allowbreak (E)\allowbreak
(A)^{-1})-\allowbreak \frac{1}{8}(\frac{1}{8}(B)^{3}\allowbreak
(A)^{-3}-\allowbreak \frac{1}{2}(B)\allowbreak (C)\allowbreak
(A)^{-2}+\allowbreak (D)\allowbreak (A)^{-1})^{2})\allowbreak +(\frac{1}{4}%
\allowbreak (-\frac{1}{108}\allowbreak (-\frac{3}{8}(B)^{2}\allowbreak
(A)^{-2}+\allowbreak (C)\allowbreak (A)^{-1})^{3}+\allowbreak \frac{1}{3}(-%
\frac{3}{8}(B)^{2}\allowbreak (A)^{-2}+\allowbreak (C)\allowbreak
(A)^{-1})\allowbreak (-\frac{3}{256}(B)^{4}\allowbreak
(A)^{-4}+\allowbreak
\frac{1}{16}(C)\allowbreak (B)^{2}\allowbreak (A)^{-3}-\allowbreak \frac{1}{4%
}(B)\allowbreak (D)\allowbreak (A)^{-2}+\allowbreak (E)\allowbreak
(A)^{-1})-\allowbreak \frac{1}{8}(\frac{1}{8}(B)^{3}\allowbreak
(A)^{-3}-\allowbreak \frac{1}{2}(B)\allowbreak (C)\allowbreak
(A)^{-2}+\allowbreak (D)\allowbreak (A)^{-1})^{2})^{2}+\allowbreak \frac{1}{%
27}\allowbreak (-\frac{1}{12}\allowbreak
(-\frac{3}{8}(B)^{2}\allowbreak
(A)^{-2}+\allowbreak (C)\allowbreak (A)^{-1})^{2}-\allowbreak (-\frac{3}{256}%
(B)^{4}\allowbreak (A)^{-4}+\allowbreak \frac{1}{16}(C)\allowbreak
(B)^{2}\allowbreak (A)^{-3}-\allowbreak \frac{1}{4}(B)\allowbreak
(D)\allowbreak (A)^{-2}+\allowbreak (E)\allowbreak (A)^{-1}))^{3})^{\frac{1}{%
2}}))^{\frac{1}{3}})^{-1}))^{\frac{1}{2}})^{-1}))^{\frac{1}{2}}).$

The second root is

$x_{2}=-\frac{1}{4}(B)\allowbreak (A)^{-1}+\allowbreak \frac{1}{2}%
\allowbreak ((((-\frac{3}{8}(B)^{2}\allowbreak (A)^{-2}+\allowbreak
(C)\allowbreak (A)^{-1})+\allowbreak 2(-\frac{5}{6}\allowbreak (-\frac{3}{8}%
(B)^{2}\allowbreak (A)^{-2}+\allowbreak (C)\allowbreak
(A)^{-1})+\allowbreak (((-\frac{1}{2}(-\frac{1}{108}\allowbreak
(-\frac{3}{8}(B)^{2}\allowbreak
(A)^{-2}+\allowbreak (C)\allowbreak (A)^{-1})^{3}+\allowbreak \frac{1}{3}(-%
\frac{3}{8}(B)^{2}\allowbreak (A)^{-2}+\allowbreak (C)\allowbreak
(A)^{-1})\allowbreak (-\frac{3}{256}(B)^{4}\allowbreak
(A)^{-4}+\allowbreak
\frac{1}{16}(C)\allowbreak (B)^{2}\allowbreak (A)^{-3}-\allowbreak \frac{1}{4%
}(B)\allowbreak (D)\allowbreak (A)^{-2}+\allowbreak (E)\allowbreak
(A)^{-1})-\allowbreak \frac{1}{8}(\frac{1}{8}(B)^{3}\allowbreak
(A)^{-3}-\allowbreak \frac{1}{2}(B)\allowbreak (C)\allowbreak
(A)^{-2}+\allowbreak (D)\allowbreak (A)^{-1})^{2})\allowbreak +(\frac{1}{4}%
\allowbreak (-\frac{1}{108}\allowbreak (-\frac{3}{8}(B)^{2}\allowbreak
(A)^{-2}+\allowbreak (C)\allowbreak (A)^{-1})^{3}+\allowbreak \frac{1}{3}(-%
\frac{3}{8}(B)^{2}\allowbreak (A)^{-2}+\allowbreak (C)\allowbreak
(A)^{-1})\allowbreak (-\frac{3}{256}(B)^{4}\allowbreak
(A)^{-4}+\allowbreak
\frac{1}{16}(C)\allowbreak (B)^{2}\allowbreak (A)^{-3}-\allowbreak \frac{1}{4%
}(B)\allowbreak (D)\allowbreak (A)^{-2}+\allowbreak (E)\allowbreak
(A)^{-1})-\allowbreak \frac{1}{8}(\frac{1}{8}(B)^{3}\allowbreak
(A)^{-3}-\allowbreak \frac{1}{2}(B)\allowbreak (C)\allowbreak
(A)^{-2}+\allowbreak (D)\allowbreak (A)^{-1})^{2})^{2}+\allowbreak \frac{1}{%
27}\allowbreak (-\frac{1}{12}\allowbreak
(-\frac{3}{8}(B)^{2}\allowbreak
(A)^{-2}+\allowbreak (C)\allowbreak (A)^{-1})^{2}-\allowbreak (-\frac{3}{256}%
(B)^{4}\allowbreak (A)^{-4}+\allowbreak \frac{1}{16}(C)\allowbreak
(B)^{2}\allowbreak (A)^{-3}-\allowbreak \frac{1}{4}(B)\allowbreak
(D)\allowbreak (A)^{-2}+\allowbreak (E)\allowbreak (A)^{-1}))^{3})^{\frac{1}{%
2}}))^{\frac{1}{3}})-\allowbreak \frac{1}{3}(-\frac{1}{12}\allowbreak (-%
\frac{3}{8}(B)^{2}\allowbreak (A)^{-2}+\allowbreak (C)\allowbreak
(A)^{-1})^{2}-\allowbreak (-\frac{3}{256}(B)^{4}\allowbreak
(A)^{-4}+\allowbreak \frac{1}{16}(C)\allowbreak (B)^{2}\allowbreak
(A)^{-3}-\allowbreak \frac{1}{4}(B)\allowbreak (D)\allowbreak
(A)^{-2}+\allowbreak (E)\allowbreak (A)^{-1}))\allowbreak (((-\frac{1}{2}(-%
\frac{1}{108}\allowbreak (-\frac{3}{8}(B)^{2}\allowbreak
(A)^{-2}+\allowbreak (C)\allowbreak (A)^{-1})^{3}+\allowbreak \frac{1}{3}(-%
\frac{3}{8}(B)^{2}\allowbreak (A)^{-2}+\allowbreak (C)\allowbreak
(A)^{-1})\allowbreak (-\frac{3}{256}(B)^{4}\allowbreak
(A)^{-4}+\allowbreak
\frac{1}{16}(C)\allowbreak (B)^{2}\allowbreak (A)^{-3}-\allowbreak \frac{1}{4%
}(B)\allowbreak (D)\allowbreak (A)^{-2}+\allowbreak (E)\allowbreak
(A)^{-1})-\allowbreak \frac{1}{8}(\frac{1}{8}(B)^{3}\allowbreak
(A)^{-3}-\allowbreak \frac{1}{2}(B)\allowbreak (C)\allowbreak
(A)^{-2}+\allowbreak (D)\allowbreak (A)^{-1})^{2})\allowbreak +(\frac{1}{4}%
\allowbreak (-\frac{1}{108}\allowbreak (-\frac{3}{8}(B)^{2}\allowbreak
(A)^{-2}+\allowbreak (C)\allowbreak (A)^{-1})^{3}+\allowbreak \frac{1}{3}(-%
\frac{3}{8}(B)^{2}\allowbreak (A)^{-2}+\allowbreak (C)\allowbreak
(A)^{-1})\allowbreak (-\frac{3}{256}(B)^{4}\allowbreak
(A)^{-4}+\allowbreak
\frac{1}{16}(C)\allowbreak (B)^{2}\allowbreak (A)^{-3}-\allowbreak \frac{1}{4%
}(B)\allowbreak (D)\allowbreak (A)^{-2}+\allowbreak (E)\allowbreak
(A)^{-1})-\allowbreak \frac{1}{8}(\frac{1}{8}(B)^{3}\allowbreak
(A)^{-3}-\allowbreak \frac{1}{2}(B)\allowbreak (C)\allowbreak
(A)^{-2}+\allowbreak (D)\allowbreak (A)^{-1})^{2})^{2}+\allowbreak \frac{1}{%
27}\allowbreak (-\frac{1}{12}\allowbreak
(-\frac{3}{8}(B)^{2}\allowbreak
(A)^{-2}+\allowbreak (C)\allowbreak (A)^{-1})^{2}-\allowbreak (-\frac{3}{256}%
(B)^{4}\allowbreak (A)^{-4}+\allowbreak \frac{1}{16}(C)\allowbreak
(B)^{2}\allowbreak (A)^{-3}-\allowbreak \frac{1}{4}(B)\allowbreak
(D)\allowbreak (A)^{-2}+\allowbreak (E)\allowbreak (A)^{-1}))^{3})^{\frac{1}{%
2}}))^{\frac{1}{3}})^{-1}))^{\frac{1}{2}})+\allowbreak (-(3\allowbreak (-%
\frac{3}{8}(B)^{2}\allowbreak (A)^{-2}+\allowbreak (C)\allowbreak
(A)^{-1})+\allowbreak 2\allowbreak (-\frac{5}{6}\allowbreak (-\frac{3}{8}%
(B)^{2}\allowbreak (A)^{-2}+\allowbreak (C)\allowbreak
(A)^{-1})+\allowbreak (((-\frac{1}{2}(-\frac{1}{108}\allowbreak
(-\frac{3}{8}(B)^{2}\allowbreak
(A)^{-2}+\allowbreak (C)\allowbreak (A)^{-1})^{3}+\allowbreak \frac{1}{3}(-%
\frac{3}{8}(B)^{2}\allowbreak (A)^{-2}+\allowbreak (C)\allowbreak
(A)^{-1})\allowbreak (-\frac{3}{256}(B)^{4}\allowbreak
(A)^{-4}+\allowbreak
\frac{1}{16}(C)\allowbreak (B)^{2}\allowbreak (A)^{-3}-\allowbreak \frac{1}{4%
}(B)\allowbreak (D)\allowbreak (A)^{-2}+\allowbreak (E)\allowbreak
(A)^{-1})-\allowbreak \frac{1}{8}(\frac{1}{8}(B)^{3}\allowbreak
(A)^{-3}-\allowbreak \frac{1}{2}(B)\allowbreak (C)\allowbreak
(A)^{-2}+\allowbreak (D)\allowbreak (A)^{-1})^{2})\allowbreak +(\frac{1}{4}%
\allowbreak (-\frac{1}{108}\allowbreak (-\frac{3}{8}(B)^{2}\allowbreak
(A)^{-2}+\allowbreak (C)\allowbreak (A)^{-1})^{3}+\allowbreak \frac{1}{3}(-%
\frac{3}{8}(B)^{2}\allowbreak (A)^{-2}+\allowbreak (C)\allowbreak
(A)^{-1})\allowbreak (-\frac{3}{256}(B)^{4}\allowbreak
(A)^{-4}+\allowbreak
\frac{1}{16}(C)\allowbreak (B)^{2}\allowbreak (A)^{-3}-\allowbreak \frac{1}{4%
}(B)\allowbreak (D)\allowbreak (A)^{-2}+\allowbreak (E)\allowbreak
(A)^{-1})-\allowbreak \frac{1}{8}(\frac{1}{8}(B)^{3}\allowbreak
(A)^{-3}-\allowbreak \frac{1}{2}(B)\allowbreak (C)\allowbreak
(A)^{-2}+\allowbreak (D)\allowbreak (A)^{-1})^{2})^{2}+\allowbreak \frac{1}{%
27}\allowbreak (-\frac{1}{12}\allowbreak
(-\frac{3}{8}(B)^{2}\allowbreak
(A)^{-2}+\allowbreak (C)\allowbreak (A)^{-1})^{2}-\allowbreak (-\frac{3}{256}%
(B)^{4}\allowbreak (A)^{-4}+\allowbreak \frac{1}{16}(C)\allowbreak
(B)^{2}\allowbreak (A)^{-3}-\allowbreak \frac{1}{4}(B)\allowbreak
(D)\allowbreak (A)^{-2}+\allowbreak (E)\allowbreak (A)^{-1}))^{3})^{\frac{1}{%
2}}))^{\frac{1}{3}})-\allowbreak \frac{1}{3}(-\frac{1}{12}\allowbreak (-%
\frac{3}{8}(B)^{2}\allowbreak (A)^{-2}+\allowbreak (C)\allowbreak
(A)^{-1})^{2}-\allowbreak (-\frac{3}{256}(B)^{4}\allowbreak
(A)^{-4}+\allowbreak \frac{1}{16}(C)\allowbreak (B)^{2}\allowbreak
(A)^{-3}-\allowbreak \frac{1}{4}(B)\allowbreak (D)\allowbreak
(A)^{-2}+\allowbreak (E)\allowbreak (A)^{-1}))\allowbreak (((-\frac{1}{2}(-%
\frac{1}{108}\allowbreak (-\frac{3}{8}(B)^{2}\allowbreak
(A)^{-2}+\allowbreak (C)\allowbreak (A)^{-1})^{3}+\allowbreak \frac{1}{3}(-%
\frac{3}{8}(B)^{2}\allowbreak (A)^{-2}+\allowbreak (C)\allowbreak
(A)^{-1})\allowbreak (-\frac{3}{256}(B)^{4}\allowbreak
(A)^{-4}+\allowbreak
\frac{1}{16}(C)\allowbreak (B)^{2}\allowbreak (A)^{-3}-\allowbreak \frac{1}{4%
}(B)\allowbreak (D)\allowbreak (A)^{-2}+\allowbreak (E)\allowbreak
(A)^{-1})-\allowbreak \frac{1}{8}(\frac{1}{8}(B)^{3}\allowbreak
(A)^{-3}-\allowbreak \frac{1}{2}(B)\allowbreak (C)\allowbreak
(A)^{-2}+\allowbreak (D)\allowbreak (A)^{-1})^{2})\allowbreak +(\frac{1}{4}%
\allowbreak (-\frac{1}{108}\allowbreak (-\frac{3}{8}(B)^{2}\allowbreak
(A)^{-2}+\allowbreak (C)\allowbreak (A)^{-1})^{3}+\allowbreak \frac{1}{3}(-%
\frac{3}{8}(B)^{2}\allowbreak (A)^{-2}+\allowbreak (C)\allowbreak
(A)^{-1})\allowbreak (-\frac{3}{256}(B)^{4}\allowbreak
(A)^{-4}+\allowbreak
\frac{1}{16}(C)\allowbreak (B)^{2}\allowbreak (A)^{-3}-\allowbreak \frac{1}{4%
}(B)\allowbreak (D)\allowbreak (A)^{-2}+\allowbreak (E)\allowbreak
(A)^{-1})-\allowbreak \frac{1}{8}(\frac{1}{8}(B)^{3}\allowbreak
(A)^{-3}-\allowbreak \frac{1}{2}(B)\allowbreak (C)\allowbreak
(A)^{-2}+\allowbreak (D)\allowbreak (A)^{-1})^{2})^{2}+\allowbreak \frac{1}{%
27}\allowbreak (-\frac{1}{12}\allowbreak
(-\frac{3}{8}(B)^{2}\allowbreak
(A)^{-2}+\allowbreak (C)\allowbreak (A)^{-1})^{2}-\allowbreak (-\frac{3}{256}%
(B)^{4}\allowbreak (A)^{-4}+\allowbreak \frac{1}{16}(C)\allowbreak
(B)^{2}\allowbreak (A)^{-3}-\allowbreak \frac{1}{4}(B)\allowbreak
(D)\allowbreak (A)^{-2}+\allowbreak (E)\allowbreak (A)^{-1}))^{3})^{\frac{1}{%
2}}))^{\frac{1}{3}})^{-1})+\allowbreak 2\allowbreak (\frac{1}{8}%
(B)^{3}\allowbreak (A)^{-3}-\allowbreak \frac{1}{2}(B)\allowbreak
(C)\allowbreak (A)^{-2}+\allowbreak (D)\allowbreak (A)^{-1})\allowbreak (((-%
\frac{3}{8}(B)^{2}\allowbreak (A)^{-2}+\allowbreak (C)\allowbreak
(A)^{-1})+\allowbreak 2(-\frac{5}{6}\allowbreak (-\frac{3}{8}%
(B)^{2}\allowbreak (A)^{-2}+\allowbreak (C)\allowbreak
(A)^{-1})+\allowbreak (((-\frac{1}{2}(-\frac{1}{108}\allowbreak
(-\frac{3}{8}(B)^{2}\allowbreak
(A)^{-2}+\allowbreak (C)\allowbreak (A)^{-1})^{3}+\allowbreak \frac{1}{3}(-%
\frac{3}{8}(B)^{2}\allowbreak (A)^{-2}+\allowbreak (C)\allowbreak
(A)^{-1})\allowbreak (-\frac{3}{256}(B)^{4}\allowbreak
(A)^{-4}+\allowbreak
\frac{1}{16}(C)\allowbreak (B)^{2}\allowbreak (A)^{-3}-\allowbreak \frac{1}{4%
}(B)\allowbreak (D)\allowbreak (A)^{-2}+\allowbreak (E)\allowbreak
(A)^{-1})-\allowbreak \frac{1}{8}(\frac{1}{8}(B)^{3}\allowbreak
(A)^{-3}-\allowbreak \frac{1}{2}(B)\allowbreak (C)\allowbreak
(A)^{-2}+\allowbreak (D)\allowbreak (A)^{-1})^{2})\allowbreak +(\frac{1}{4}%
\allowbreak (-\frac{1}{108}\allowbreak (-\frac{3}{8}(B)^{2}\allowbreak
(A)^{-2}+\allowbreak (C)\allowbreak (A)^{-1})^{3}+\allowbreak \frac{1}{3}(-%
\frac{3}{8}(B)^{2}\allowbreak (A)^{-2}+\allowbreak (C)\allowbreak
(A)^{-1})\allowbreak (-\frac{3}{256}(B)^{4}\allowbreak
(A)^{-4}+\allowbreak
\frac{1}{16}(C)\allowbreak (B)^{2}\allowbreak (A)^{-3}-\allowbreak \frac{1}{4%
}(B)\allowbreak (D)\allowbreak (A)^{-2}+\allowbreak (E)\allowbreak
(A)^{-1})-\allowbreak \frac{1}{8}(\frac{1}{8}(B)^{3}\allowbreak
(A)^{-3}-\allowbreak \frac{1}{2}(B)\allowbreak (C)\allowbreak
(A)^{-2}+\allowbreak (D)\allowbreak (A)^{-1})^{2})^{2}+\allowbreak \frac{1}{%
27}\allowbreak (-\frac{1}{12}\allowbreak
(-\frac{3}{8}(B)^{2}\allowbreak
(A)^{-2}+\allowbreak (C)\allowbreak (A)^{-1})^{2}-\allowbreak (-\frac{3}{256}%
(B)^{4}\allowbreak (A)^{-4}+\allowbreak \frac{1}{16}(C)\allowbreak
(B)^{2}\allowbreak (A)^{-3}-\allowbreak \frac{1}{4}(B)\allowbreak
(D)\allowbreak (A)^{-2}+\allowbreak (E)\allowbreak (A)^{-1}))^{3})^{\frac{1}{%
2}}))^{\frac{1}{3}})-\allowbreak \frac{1}{3}(-\frac{1}{12}\allowbreak (-%
\frac{3}{8}(B)^{2}\allowbreak (A)^{-2}+\allowbreak (C)\allowbreak
(A)^{-1})^{2}-\allowbreak (-\frac{3}{256}(B)^{4}\allowbreak
(A)^{-4}+\allowbreak \frac{1}{16}(C)\allowbreak (B)^{2}\allowbreak
(A)^{-3}-\allowbreak \frac{1}{4}(B)\allowbreak (D)\allowbreak
(A)^{-2}+\allowbreak (E)\allowbreak (A)^{-1}))\allowbreak (((-\frac{1}{2}(-%
\frac{1}{108}\allowbreak (-\frac{3}{8}(B)^{2}\allowbreak
(A)^{-2}+\allowbreak (C)\allowbreak (A)^{-1})^{3}+\allowbreak \frac{1}{3}(-%
\frac{3}{8}(B)^{2}\allowbreak (A)^{-2}+\allowbreak (C)\allowbreak
(A)^{-1})\allowbreak (-\frac{3}{256}(B)^{4}\allowbreak
(A)^{-4}+\allowbreak
\frac{1}{16}(C)\allowbreak (B)^{2}\allowbreak (A)^{-3}-\allowbreak \frac{1}{4%
}(B)\allowbreak (D)\allowbreak (A)^{-2}+\allowbreak (E)\allowbreak
(A)^{-1})-\allowbreak \frac{1}{8}(\frac{1}{8}(B)^{3}\allowbreak
(A)^{-3}-\allowbreak \frac{1}{2}(B)\allowbreak (C)\allowbreak
(A)^{-2}+\allowbreak (D)\allowbreak (A)^{-1})^{2})\allowbreak +(\frac{1}{4}%
\allowbreak (-\frac{1}{108}\allowbreak (-\frac{3}{8}(B)^{2}\allowbreak
(A)^{-2}+\allowbreak (C)\allowbreak (A)^{-1})^{3}+\allowbreak \frac{1}{3}(-%
\frac{3}{8}(B)^{2}\allowbreak (A)^{-2}+\allowbreak (C)\allowbreak
(A)^{-1})\allowbreak (-\frac{3}{256}(B)^{4}\allowbreak
(A)^{-4}+\allowbreak
\frac{1}{16}(C)\allowbreak (B)^{2}\allowbreak (A)^{-3}-\allowbreak \frac{1}{4%
}(B)\allowbreak (D)\allowbreak (A)^{-2}+\allowbreak (E)\allowbreak
(A)^{-1})-\allowbreak \frac{1}{8}(\frac{1}{8}(B)^{3}\allowbreak
(A)^{-3}-\allowbreak \frac{1}{2}(B)\allowbreak (C)\allowbreak
(A)^{-2}+\allowbreak (D)\allowbreak (A)^{-1})^{2})^{2}+\allowbreak \frac{1}{%
27}\allowbreak (-\frac{1}{12}\allowbreak
(-\frac{3}{8}(B)^{2}\allowbreak
(A)^{-2}+\allowbreak (C)\allowbreak (A)^{-1})^{2}-\allowbreak (-\frac{3}{256}%
(B)^{4}\allowbreak (A)^{-4}+\allowbreak \frac{1}{16}(C)\allowbreak
(B)^{2}\allowbreak (A)^{-3}-\allowbreak \frac{1}{4}(B)\allowbreak
(D)\allowbreak (A)^{-2}+\allowbreak (E)\allowbreak (A)^{-1}))^{3})^{\frac{1}{%
2}}))^{\frac{1}{3}})^{-1}))^{\frac{1}{2}})^{-1}))^{\frac{1}{2}}).$

The third root is

$x_{3}=-\frac{1}{4}(B)\allowbreak (A)^{-1}+\allowbreak \frac{1}{2}%
\allowbreak (-(((-\frac{3}{8}(B)^{2}\allowbreak (A)^{-2}+\allowbreak
(C)\allowbreak (A)^{-1})+\allowbreak 2(-\frac{5}{6}\allowbreak (-\frac{3}{8}%
(B)^{2}\allowbreak (A)^{-2}+\allowbreak (C)\allowbreak
(A)^{-1})+\allowbreak (((-\frac{1}{2}(-\frac{1}{108}\allowbreak
(-\frac{3}{8}(B)^{2}\allowbreak
(A)^{-2}+\allowbreak (C)\allowbreak (A)^{-1})^{3}+\allowbreak \frac{1}{3}(-%
\frac{3}{8}(B)^{2}\allowbreak (A)^{-2}+\allowbreak (C)\allowbreak
(A)^{-1})\allowbreak (-\frac{3}{256}(B)^{4}\allowbreak
(A)^{-4}+\allowbreak
\frac{1}{16}(C)\allowbreak (B)^{2}\allowbreak (A)^{-3}-\allowbreak \frac{1}{4%
}(B)\allowbreak (D)\allowbreak (A)^{-2}+\allowbreak (E)\allowbreak
(A)^{-1})-\allowbreak \frac{1}{8}(\frac{1}{8}(B)^{3}\allowbreak
(A)^{-3}-\allowbreak \frac{1}{2}(B)\allowbreak (C)\allowbreak
(A)^{-2}+\allowbreak (D)\allowbreak (A)^{-1})^{2})\allowbreak +(\frac{1}{4}%
\allowbreak (-\frac{1}{108}\allowbreak (-\frac{3}{8}(B)^{2}\allowbreak
(A)^{-2}+\allowbreak (C)\allowbreak (A)^{-1})^{3}+\allowbreak \frac{1}{3}(-%
\frac{3}{8}(B)^{2}\allowbreak (A)^{-2}+\allowbreak (C)\allowbreak
(A)^{-1})\allowbreak (-\frac{3}{256}(B)^{4}\allowbreak
(A)^{-4}+\allowbreak
\frac{1}{16}(C)\allowbreak (B)^{2}\allowbreak (A)^{-3}-\allowbreak \frac{1}{4%
}(B)\allowbreak (D)\allowbreak (A)^{-2}+\allowbreak (E)\allowbreak
(A)^{-1})-\allowbreak \frac{1}{8}(\frac{1}{8}(B)^{3}\allowbreak
(A)^{-3}-\allowbreak \frac{1}{2}(B)\allowbreak (C)\allowbreak
(A)^{-2}+\allowbreak (D)\allowbreak (A)^{-1})^{2})^{2}+\allowbreak \frac{1}{%
27}\allowbreak (-\frac{1}{12}\allowbreak
(-\frac{3}{8}(B)^{2}\allowbreak
(A)^{-2}+\allowbreak (C)\allowbreak (A)^{-1})^{2}-\allowbreak (-\frac{3}{256}%
(B)^{4}\allowbreak (A)^{-4}+\allowbreak \frac{1}{16}(C)\allowbreak
(B)^{2}\allowbreak (A)^{-3}-\allowbreak \frac{1}{4}(B)\allowbreak
(D)\allowbreak (A)^{-2}+\allowbreak (E)\allowbreak (A)^{-1}))^{3})^{\frac{1}{%
2}}))^{\frac{1}{3}})-\allowbreak \frac{1}{3}(-\frac{1}{12}\allowbreak (-%
\frac{3}{8}(B)^{2}\allowbreak (A)^{-2}+\allowbreak (C)\allowbreak
(A)^{-1})^{2}-\allowbreak (-\frac{3}{256}(B)^{4}\allowbreak
(A)^{-4}+\allowbreak \frac{1}{16}(C)\allowbreak (B)^{2}\allowbreak
(A)^{-3}-\allowbreak \frac{1}{4}(B)\allowbreak (D)\allowbreak
(A)^{-2}+\allowbreak (E)\allowbreak (A)^{-1}))\allowbreak (((-\frac{1}{2}(-%
\frac{1}{108}\allowbreak (-\frac{3}{8}(B)^{2}\allowbreak
(A)^{-2}+\allowbreak (C)\allowbreak (A)^{-1})^{3}+\allowbreak \frac{1}{3}(-%
\frac{3}{8}(B)^{2}\allowbreak (A)^{-2}+\allowbreak (C)\allowbreak
(A)^{-1})\allowbreak (-\frac{3}{256}(B)^{4}\allowbreak
(A)^{-4}+\allowbreak
\frac{1}{16}(C)\allowbreak (B)^{2}\allowbreak (A)^{-3}-\allowbreak \frac{1}{4%
}(B)\allowbreak (D)\allowbreak (A)^{-2}+\allowbreak (E)\allowbreak
(A)^{-1})-\allowbreak \frac{1}{8}(\frac{1}{8}(B)^{3}\allowbreak
(A)^{-3}-\allowbreak \frac{1}{2}(B)\allowbreak (C)\allowbreak
(A)^{-2}+\allowbreak (D)\allowbreak (A)^{-1})^{2})\allowbreak +(\frac{1}{4}%
\allowbreak (-\frac{1}{108}\allowbreak (-\frac{3}{8}(B)^{2}\allowbreak
(A)^{-2}+\allowbreak (C)\allowbreak (A)^{-1})^{3}+\allowbreak \frac{1}{3}(-%
\frac{3}{8}(B)^{2}\allowbreak (A)^{-2}+\allowbreak (C)\allowbreak
(A)^{-1})\allowbreak (-\frac{3}{256}(B)^{4}\allowbreak
(A)^{-4}+\allowbreak
\frac{1}{16}(C)\allowbreak (B)^{2}\allowbreak (A)^{-3}-\allowbreak \frac{1}{4%
}(B)\allowbreak (D)\allowbreak (A)^{-2}+\allowbreak (E)\allowbreak
(A)^{-1})-\allowbreak \frac{1}{8}(\frac{1}{8}(B)^{3}\allowbreak
(A)^{-3}-\allowbreak \frac{1}{2}(B)\allowbreak (C)\allowbreak
(A)^{-2}+\allowbreak (D)\allowbreak (A)^{-1})^{2})^{2}+\allowbreak \frac{1}{%
27}\allowbreak (-\frac{1}{12}\allowbreak
(-\frac{3}{8}(B)^{2}\allowbreak
(A)^{-2}+\allowbreak (C)\allowbreak (A)^{-1})^{2}-\allowbreak (-\frac{3}{256}%
(B)^{4}\allowbreak (A)^{-4}+\allowbreak \frac{1}{16}(C)\allowbreak
(B)^{2}\allowbreak (A)^{-3}-\allowbreak \frac{1}{4}(B)\allowbreak
(D)\allowbreak (A)^{-2}+\allowbreak (E)\allowbreak (A)^{-1}))^{3})^{\frac{1}{%
2}}))^{\frac{1}{3}})^{-1}))^{\frac{1}{2}})-\allowbreak (-(3\allowbreak (-%
\frac{3}{8}(B)^{2}\allowbreak (A)^{-2}+\allowbreak (C)\allowbreak
(A)^{-1})+\allowbreak 2\allowbreak (-\frac{5}{6}\allowbreak (-\frac{3}{8}%
(B)^{2}\allowbreak (A)^{-2}+\allowbreak (C)\allowbreak
(A)^{-1})+\allowbreak (((-\frac{1}{2}(-\frac{1}{108}\allowbreak
(-\frac{3}{8}(B)^{2}\allowbreak
(A)^{-2}+\allowbreak (C)\allowbreak (A)^{-1})^{3}+\allowbreak \frac{1}{3}(-%
\frac{3}{8}(B)^{2}\allowbreak (A)^{-2}+\allowbreak (C)\allowbreak
(A)^{-1})\allowbreak (-\frac{3}{256}(B)^{4}\allowbreak
(A)^{-4}+\allowbreak
\frac{1}{16}(C)\allowbreak (B)^{2}\allowbreak (A)^{-3}-\allowbreak \frac{1}{4%
}(B)\allowbreak (D)\allowbreak (A)^{-2}+\allowbreak (E)\allowbreak
(A)^{-1})-\allowbreak \frac{1}{8}(\frac{1}{8}(B)^{3}\allowbreak
(A)^{-3}-\allowbreak \frac{1}{2}(B)\allowbreak (C)\allowbreak
(A)^{-2}+\allowbreak (D)\allowbreak (A)^{-1})^{2})\allowbreak +(\frac{1}{4}%
\allowbreak (-\frac{1}{108}\allowbreak (-\frac{3}{8}(B)^{2}\allowbreak
(A)^{-2}+\allowbreak (C)\allowbreak (A)^{-1})^{3}+\allowbreak \frac{1}{3}(-%
\frac{3}{8}(B)^{2}\allowbreak (A)^{-2}+\allowbreak (C)\allowbreak
(A)^{-1})\allowbreak (-\frac{3}{256}(B)^{4}\allowbreak
(A)^{-4}+\allowbreak
\frac{1}{16}(C)\allowbreak (B)^{2}\allowbreak (A)^{-3}-\allowbreak \frac{1}{4%
}(B)\allowbreak (D)\allowbreak (A)^{-2}+\allowbreak (E)\allowbreak
(A)^{-1})-\allowbreak \frac{1}{8}(\frac{1}{8}(B)^{3}\allowbreak
(A)^{-3}-\allowbreak \frac{1}{2}(B)\allowbreak (C)\allowbreak
(A)^{-2}+\allowbreak (D)\allowbreak (A)^{-1})^{2})^{2}+\allowbreak \frac{1}{%
27}\allowbreak (-\frac{1}{12}\allowbreak
(-\frac{3}{8}(B)^{2}\allowbreak
(A)^{-2}+\allowbreak (C)\allowbreak (A)^{-1})^{2}-\allowbreak (-\frac{3}{256}%
(B)^{4}\allowbreak (A)^{-4}+\allowbreak \frac{1}{16}(C)\allowbreak
(B)^{2}\allowbreak (A)^{-3}-\allowbreak \frac{1}{4}(B)\allowbreak
(D)\allowbreak (A)^{-2}+\allowbreak (E)\allowbreak (A)^{-1}))^{3})^{\frac{1}{%
2}}))^{\frac{1}{3}})-\allowbreak \frac{1}{3}(-\frac{1}{12}\allowbreak (-%
\frac{3}{8}(B)^{2}\allowbreak (A)^{-2}+\allowbreak (C)\allowbreak
(A)^{-1})^{2}-\allowbreak (-\frac{3}{256}(B)^{4}\allowbreak
(A)^{-4}+\allowbreak \frac{1}{16}(C)\allowbreak (B)^{2}\allowbreak
(A)^{-3}-\allowbreak \frac{1}{4}(B)\allowbreak (D)\allowbreak
(A)^{-2}+\allowbreak (E)\allowbreak (A)^{-1}))\allowbreak (((-\frac{1}{2}(-%
\frac{1}{108}\allowbreak (-\frac{3}{8}(B)^{2}\allowbreak
(A)^{-2}+\allowbreak (C)\allowbreak (A)^{-1})^{3}+\allowbreak \frac{1}{3}(-%
\frac{3}{8}(B)^{2}\allowbreak (A)^{-2}+\allowbreak (C)\allowbreak
(A)^{-1})\allowbreak (-\frac{3}{256}(B)^{4}\allowbreak
(A)^{-4}+\allowbreak
\frac{1}{16}(C)\allowbreak (B)^{2}\allowbreak (A)^{-3}-\allowbreak \frac{1}{4%
}(B)\allowbreak (D)\allowbreak (A)^{-2}+\allowbreak (E)\allowbreak
(A)^{-1})-\allowbreak \frac{1}{8}(\frac{1}{8}(B)^{3}\allowbreak
(A)^{-3}-\allowbreak \frac{1}{2}(B)\allowbreak (C)\allowbreak
(A)^{-2}+\allowbreak (D)\allowbreak (A)^{-1})^{2})\allowbreak +(\frac{1}{4}%
\allowbreak (-\frac{1}{108}\allowbreak (-\frac{3}{8}(B)^{2}\allowbreak
(A)^{-2}+\allowbreak (C)\allowbreak (A)^{-1})^{3}+\allowbreak \frac{1}{3}(-%
\frac{3}{8}(B)^{2}\allowbreak (A)^{-2}+\allowbreak (C)\allowbreak
(A)^{-1})\allowbreak (-\frac{3}{256}(B)^{4}\allowbreak
(A)^{-4}+\allowbreak
\frac{1}{16}(C)\allowbreak (B)^{2}\allowbreak (A)^{-3}-\allowbreak \frac{1}{4%
}(B)\allowbreak (D)\allowbreak (A)^{-2}+\allowbreak (E)\allowbreak
(A)^{-1})-\allowbreak \frac{1}{8}(\frac{1}{8}(B)^{3}\allowbreak
(A)^{-3}-\allowbreak \frac{1}{2}(B)\allowbreak (C)\allowbreak
(A)^{-2}+\allowbreak (D)\allowbreak (A)^{-1})^{2})^{2}+\allowbreak \frac{1}{%
27}\allowbreak (-\frac{1}{12}\allowbreak
(-\frac{3}{8}(B)^{2}\allowbreak
(A)^{-2}+\allowbreak (C)\allowbreak (A)^{-1})^{2}-\allowbreak (-\frac{3}{256}%
(B)^{4}\allowbreak (A)^{-4}+\allowbreak \frac{1}{16}(C)\allowbreak
(B)^{2}\allowbreak (A)^{-3}-\allowbreak \frac{1}{4}(B)\allowbreak
(D)\allowbreak (A)^{-2}+\allowbreak (E)\allowbreak (A)^{-1}))^{3})^{\frac{1}{%
2}}))^{\frac{1}{3}})^{-1})-\allowbreak 2\allowbreak (\frac{1}{8}%
(B)^{3}\allowbreak (A)^{-3}-\allowbreak \frac{1}{2}(B)\allowbreak
(C)\allowbreak (A)^{-2}+\allowbreak (D)\allowbreak (A)^{-1})\allowbreak (((-%
\frac{3}{8}(B)^{2}\allowbreak (A)^{-2}+\allowbreak (C)\allowbreak
(A)^{-1})+\allowbreak 2(-\frac{5}{6}\allowbreak (-\frac{3}{8}%
(B)^{2}\allowbreak (A)^{-2}+\allowbreak (C)\allowbreak
(A)^{-1})+\allowbreak (((-\frac{1}{2}(-\frac{1}{108}\allowbreak
(-\frac{3}{8}(B)^{2}\allowbreak
(A)^{-2}+\allowbreak (C)\allowbreak (A)^{-1})^{3}+\allowbreak \frac{1}{3}(-%
\frac{3}{8}(B)^{2}\allowbreak (A)^{-2}+\allowbreak (C)\allowbreak
(A)^{-1})\allowbreak (-\frac{3}{256}(B)^{4}\allowbreak
(A)^{-4}+\allowbreak
\frac{1}{16}(C)\allowbreak (B)^{2}\allowbreak (A)^{-3}-\allowbreak \frac{1}{4%
}(B)\allowbreak (D)\allowbreak (A)^{-2}+\allowbreak (E)\allowbreak
(A)^{-1})-\allowbreak \frac{1}{8}(\frac{1}{8}(B)^{3}\allowbreak
(A)^{-3}-\allowbreak \frac{1}{2}(B)\allowbreak (C)\allowbreak
(A)^{-2}+\allowbreak (D)\allowbreak (A)^{-1})^{2})\allowbreak +(\frac{1}{4}%
\allowbreak (-\frac{1}{108}\allowbreak (-\frac{3}{8}(B)^{2}\allowbreak
(A)^{-2}+\allowbreak (C)\allowbreak (A)^{-1})^{3}+\allowbreak \frac{1}{3}(-%
\frac{3}{8}(B)^{2}\allowbreak (A)^{-2}+\allowbreak (C)\allowbreak
(A)^{-1})\allowbreak (-\frac{3}{256}(B)^{4}\allowbreak
(A)^{-4}+\allowbreak
\frac{1}{16}(C)\allowbreak (B)^{2}\allowbreak (A)^{-3}-\allowbreak \frac{1}{4%
}(B)\allowbreak (D)\allowbreak (A)^{-2}+\allowbreak (E)\allowbreak
(A)^{-1})-\allowbreak \frac{1}{8}(\frac{1}{8}(B)^{3}\allowbreak
(A)^{-3}-\allowbreak \frac{1}{2}(B)\allowbreak (C)\allowbreak
(A)^{-2}+\allowbreak (D)\allowbreak (A)^{-1})^{2})^{2}+\allowbreak \frac{1}{%
27}\allowbreak (-\frac{1}{12}\allowbreak
(-\frac{3}{8}(B)^{2}\allowbreak
(A)^{-2}+\allowbreak (C)\allowbreak (A)^{-1})^{2}-\allowbreak (-\frac{3}{256}%
(B)^{4}\allowbreak (A)^{-4}+\allowbreak \frac{1}{16}(C)\allowbreak
(B)^{2}\allowbreak (A)^{-3}-\allowbreak \frac{1}{4}(B)\allowbreak
(D)\allowbreak (A)^{-2}+\allowbreak (E)\allowbreak (A)^{-1}))^{3})^{\frac{1}{%
2}}))^{\frac{1}{3}})-\allowbreak \frac{1}{3}(-\frac{1}{12}\allowbreak (-%
\frac{3}{8}(B)^{2}\allowbreak (A)^{-2}+\allowbreak (C)\allowbreak
(A)^{-1})^{2}-\allowbreak (-\frac{3}{256}(B)^{4}\allowbreak
(A)^{-4}+\allowbreak \frac{1}{16}(C)\allowbreak (B)^{2}\allowbreak
(A)^{-3}-\allowbreak \frac{1}{4}(B)\allowbreak (D)\allowbreak
(A)^{-2}+\allowbreak (E)\allowbreak (A)^{-1}))\allowbreak (((-\frac{1}{2}(-%
\frac{1}{108}\allowbreak (-\frac{3}{8}(B)^{2}\allowbreak
(A)^{-2}+\allowbreak (C)\allowbreak (A)^{-1})^{3}+\allowbreak \frac{1}{3}(-%
\frac{3}{8}(B)^{2}\allowbreak (A)^{-2}+\allowbreak (C)\allowbreak
(A)^{-1})\allowbreak (-\frac{3}{256}(B)^{4}\allowbreak
(A)^{-4}+\allowbreak
\frac{1}{16}(C)\allowbreak (B)^{2}\allowbreak (A)^{-3}-\allowbreak \frac{1}{4%
}(B)\allowbreak (D)\allowbreak (A)^{-2}+\allowbreak (E)\allowbreak
(A)^{-1})-\allowbreak \frac{1}{8}(\frac{1}{8}(B)^{3}\allowbreak
(A)^{-3}-\allowbreak \frac{1}{2}(B)\allowbreak (C)\allowbreak
(A)^{-2}+\allowbreak (D)\allowbreak (A)^{-1})^{2})\allowbreak +(\frac{1}{4}%
\allowbreak (-\frac{1}{108}\allowbreak (-\frac{3}{8}(B)^{2}\allowbreak
(A)^{-2}+\allowbreak (C)\allowbreak (A)^{-1})^{3}+\allowbreak \frac{1}{3}(-%
\frac{3}{8}(B)^{2}\allowbreak (A)^{-2}+\allowbreak (C)\allowbreak
(A)^{-1})\allowbreak (-\frac{3}{256}(B)^{4}\allowbreak
(A)^{-4}+\allowbreak
\frac{1}{16}(C)\allowbreak (B)^{2}\allowbreak (A)^{-3}-\allowbreak \frac{1}{4%
}(B)\allowbreak (D)\allowbreak (A)^{-2}+\allowbreak (E)\allowbreak
(A)^{-1})-\allowbreak \frac{1}{8}(\frac{1}{8}(B)^{3}\allowbreak
(A)^{-3}-\allowbreak \frac{1}{2}(B)\allowbreak (C)\allowbreak
(A)^{-2}+\allowbreak (D)\allowbreak (A)^{-1})^{2})^{2}+\allowbreak \frac{1}{%
27}\allowbreak (-\frac{1}{12}\allowbreak
(-\frac{3}{8}(B)^{2}\allowbreak
(A)^{-2}+\allowbreak (C)\allowbreak (A)^{-1})^{2}-\allowbreak (-\frac{3}{256}%
(B)^{4}\allowbreak (A)^{-4}+\allowbreak \frac{1}{16}(C)\allowbreak
(B)^{2}\allowbreak (A)^{-3}-\allowbreak \frac{1}{4}(B)\allowbreak
(D)\allowbreak (A)^{-2}+\allowbreak (E)\allowbreak (A)^{-1}))^{3})^{\frac{1}{%
2}}))^{\frac{1}{3}})^{-1}))^{\frac{1}{2}})^{-1}))^{\frac{1}{2}}).$

Finally, the fourth root is

$x_{4}=-\frac{1}{4}(B)\allowbreak (A)^{-1}+\allowbreak \frac{1}{2}%
\allowbreak (-(((-\frac{3}{8}(B)^{2}\allowbreak (A)^{-2}+\allowbreak
(C)\allowbreak (A)^{-1})+\allowbreak 2(-\frac{5}{6}\allowbreak (-\frac{3}{8}%
(B)^{2}\allowbreak (A)^{-2}+\allowbreak (C)\allowbreak
(A)^{-1})+\allowbreak (((-\frac{1}{2}(-\frac{1}{108}\allowbreak
(-\frac{3}{8}(B)^{2}\allowbreak
(A)^{-2}+\allowbreak (C)\allowbreak (A)^{-1})^{3}+\allowbreak \frac{1}{3}(-%
\frac{3}{8}(B)^{2}\allowbreak (A)^{-2}+\allowbreak (C)\allowbreak
(A)^{-1})\allowbreak (-\frac{3}{256}(B)^{4}\allowbreak
(A)^{-4}+\allowbreak
\frac{1}{16}(C)\allowbreak (B)^{2}\allowbreak (A)^{-3}-\allowbreak \frac{1}{4%
}(B)\allowbreak (D)\allowbreak (A)^{-2}+\allowbreak (E)\allowbreak
(A)^{-1})-\allowbreak \frac{1}{8}(\frac{1}{8}(B)^{3}\allowbreak
(A)^{-3}-\allowbreak \frac{1}{2}(B)\allowbreak (C)\allowbreak
(A)^{-2}+\allowbreak (D)\allowbreak (A)^{-1})^{2})\allowbreak +(\frac{1}{4}%
\allowbreak (-\frac{1}{108}\allowbreak (-\frac{3}{8}(B)^{2}\allowbreak
(A)^{-2}+\allowbreak (C)\allowbreak (A)^{-1})^{3}+\allowbreak \frac{1}{3}(-%
\frac{3}{8}(B)^{2}\allowbreak (A)^{-2}+\allowbreak (C)\allowbreak
(A)^{-1})\allowbreak (-\frac{3}{256}(B)^{4}\allowbreak
(A)^{-4}+\allowbreak
\frac{1}{16}(C)\allowbreak (B)^{2}\allowbreak (A)^{-3}-\allowbreak \frac{1}{4%
}(B)\allowbreak (D)\allowbreak (A)^{-2}+\allowbreak (E)\allowbreak
(A)^{-1})-\allowbreak \frac{1}{8}(\frac{1}{8}(B)^{3}\allowbreak
(A)^{-3}-\allowbreak \frac{1}{2}(B)\allowbreak (C)\allowbreak
(A)^{-2}+\allowbreak (D)\allowbreak (A)^{-1})^{2})^{2}+\allowbreak \frac{1}{%
27}\allowbreak (-\frac{1}{12}\allowbreak
(-\frac{3}{8}(B)^{2}\allowbreak
(A)^{-2}+\allowbreak (C)\allowbreak (A)^{-1})^{2}-\allowbreak (-\frac{3}{256}%
(B)^{4}\allowbreak (A)^{-4}+\allowbreak \frac{1}{16}(C)\allowbreak
(B)^{2}\allowbreak (A)^{-3}-\allowbreak \frac{1}{4}(B)\allowbreak
(D)\allowbreak (A)^{-2}+\allowbreak (E)\allowbreak (A)^{-1}))^{3})^{\frac{1}{%
2}}))^{\frac{1}{3}})-\allowbreak \frac{1}{3}(-\frac{1}{12}\allowbreak (-%
\frac{3}{8}(B)^{2}\allowbreak (A)^{-2}+\allowbreak (C)\allowbreak
(A)^{-1})^{2}-\allowbreak (-\frac{3}{256}(B)^{4}\allowbreak
(A)^{-4}+\allowbreak \frac{1}{16}(C)\allowbreak (B)^{2}\allowbreak
(A)^{-3}-\allowbreak \frac{1}{4}(B)\allowbreak (D)\allowbreak
(A)^{-2}+\allowbreak (E)\allowbreak (A)^{-1}))\allowbreak (((-\frac{1}{2}(-%
\frac{1}{108}\allowbreak (-\frac{3}{8}(B)^{2}\allowbreak
(A)^{-2}+\allowbreak (C)\allowbreak (A)^{-1})^{3}+\allowbreak \frac{1}{3}(-%
\frac{3}{8}(B)^{2}\allowbreak (A)^{-2}+\allowbreak (C)\allowbreak
(A)^{-1})\allowbreak (-\frac{3}{256}(B)^{4}\allowbreak
(A)^{-4}+\allowbreak
\frac{1}{16}(C)\allowbreak (B)^{2}\allowbreak (A)^{-3}-\allowbreak \frac{1}{4%
}(B)\allowbreak (D)\allowbreak (A)^{-2}+\allowbreak (E)\allowbreak
(A)^{-1})-\allowbreak \frac{1}{8}(\frac{1}{8}(B)^{3}\allowbreak
(A)^{-3}-\allowbreak \frac{1}{2}(B)\allowbreak (C)\allowbreak
(A)^{-2}+\allowbreak (D)\allowbreak (A)^{-1})^{2})\allowbreak +(\frac{1}{4}%
\allowbreak (-\frac{1}{108}\allowbreak (-\frac{3}{8}(B)^{2}\allowbreak
(A)^{-2}+\allowbreak (C)\allowbreak (A)^{-1})^{3}+\allowbreak \frac{1}{3}(-%
\frac{3}{8}(B)^{2}\allowbreak (A)^{-2}+\allowbreak (C)\allowbreak
(A)^{-1})\allowbreak (-\frac{3}{256}(B)^{4}\allowbreak
(A)^{-4}+\allowbreak
\frac{1}{16}(C)\allowbreak (B)^{2}\allowbreak (A)^{-3}-\allowbreak \frac{1}{4%
}(B)\allowbreak (D)\allowbreak (A)^{-2}+\allowbreak (E)\allowbreak
(A)^{-1})-\allowbreak \frac{1}{8}(\frac{1}{8}(B)^{3}\allowbreak
(A)^{-3}-\allowbreak \frac{1}{2}(B)\allowbreak (C)\allowbreak
(A)^{-2}+\allowbreak (D)\allowbreak (A)^{-1})^{2})^{2}+\allowbreak \frac{1}{%
27}\allowbreak (-\frac{1}{12}\allowbreak
(-\frac{3}{8}(B)^{2}\allowbreak
(A)^{-2}+\allowbreak (C)\allowbreak (A)^{-1})^{2}-\allowbreak (-\frac{3}{256}%
(B)^{4}\allowbreak (A)^{-4}+\allowbreak \frac{1}{16}(C)\allowbreak
(B)^{2}\allowbreak (A)^{-3}-\allowbreak \frac{1}{4}(B)\allowbreak
(D)\allowbreak (A)^{-2}+\allowbreak (E)\allowbreak (A)^{-1}))^{3})^{\frac{1}{%
2}}))^{\frac{1}{3}})^{-1}))^{\frac{1}{2}})+\allowbreak (-(3\allowbreak (-%
\frac{3}{8}(B)^{2}\allowbreak (A)^{-2}+\allowbreak (C)\allowbreak
(A)^{-1})+\allowbreak 2\allowbreak (-\frac{5}{6}\allowbreak (-\frac{3}{8}%
(B)^{2}\allowbreak (A)^{-2}+\allowbreak (C)\allowbreak
(A)^{-1})+\allowbreak (((-\frac{1}{2}(-\frac{1}{108}\allowbreak
(-\frac{3}{8}(B)^{2}\allowbreak
(A)^{-2}+\allowbreak (C)\allowbreak (A)^{-1})^{3}+\allowbreak \frac{1}{3}(-%
\frac{3}{8}(B)^{2}\allowbreak (A)^{-2}+\allowbreak (C)\allowbreak
(A)^{-1})\allowbreak (-\frac{3}{256}(B)^{4}\allowbreak
(A)^{-4}+\allowbreak
\frac{1}{16}(C)\allowbreak (B)^{2}\allowbreak (A)^{-3}-\allowbreak \frac{1}{4%
}(B)\allowbreak (D)\allowbreak (A)^{-2}+\allowbreak (E)\allowbreak
(A)^{-1})-\allowbreak \frac{1}{8}(\frac{1}{8}(B)^{3}\allowbreak
(A)^{-3}-\allowbreak \frac{1}{2}(B)\allowbreak (C)\allowbreak
(A)^{-2}+\allowbreak (D)\allowbreak (A)^{-1})^{2})\allowbreak +(\frac{1}{4}%
\allowbreak (-\frac{1}{108}\allowbreak (-\frac{3}{8}(B)^{2}\allowbreak
(A)^{-2}+\allowbreak (C)\allowbreak (A)^{-1})^{3}+\allowbreak \frac{1}{3}(-%
\frac{3}{8}(B)^{2}\allowbreak (A)^{-2}+\allowbreak (C)\allowbreak
(A)^{-1})\allowbreak (-\frac{3}{256}(B)^{4}\allowbreak
(A)^{-4}+\allowbreak
\frac{1}{16}(C)\allowbreak (B)^{2}\allowbreak (A)^{-3}-\allowbreak \frac{1}{4%
}(B)\allowbreak (D)\allowbreak (A)^{-2}+\allowbreak (E)\allowbreak
(A)^{-1})-\allowbreak \frac{1}{8}(\frac{1}{8}(B)^{3}\allowbreak
(A)^{-3}-\allowbreak \frac{1}{2}(B)\allowbreak (C)\allowbreak
(A)^{-2}+\allowbreak (D)\allowbreak (A)^{-1})^{2})^{2}+\allowbreak \frac{1}{%
27}\allowbreak (-\frac{1}{12}\allowbreak
(-\frac{3}{8}(B)^{2}\allowbreak
(A)^{-2}+\allowbreak (C)\allowbreak (A)^{-1})^{2}-\allowbreak (-\frac{3}{256}%
(B)^{4}\allowbreak (A)^{-4}+\allowbreak \frac{1}{16}(C)\allowbreak
(B)^{2}\allowbreak (A)^{-3}-\allowbreak \frac{1}{4}(B)\allowbreak
(D)\allowbreak (A)^{-2}+\allowbreak (E)\allowbreak (A)^{-1}))^{3})^{\frac{1}{%
2}}))^{\frac{1}{3}})-\allowbreak \frac{1}{3}(-\frac{1}{12}\allowbreak (-%
\frac{3}{8}(B)^{2}\allowbreak (A)^{-2}+\allowbreak (C)\allowbreak
(A)^{-1})^{2}-\allowbreak (-\frac{3}{256}(B)^{4}\allowbreak
(A)^{-4}+\allowbreak \frac{1}{16}(C)\allowbreak (B)^{2}\allowbreak
(A)^{-3}-\allowbreak \frac{1}{4}(B)\allowbreak (D)\allowbreak
(A)^{-2}+\allowbreak (E)\allowbreak (A)^{-1}))\allowbreak (((-\frac{1}{2}(-%
\frac{1}{108}\allowbreak (-\frac{3}{8}(B)^{2}\allowbreak
(A)^{-2}+\allowbreak (C)\allowbreak (A)^{-1})^{3}+\allowbreak \frac{1}{3}(-%
\frac{3}{8}(B)^{2}\allowbreak (A)^{-2}+\allowbreak (C)\allowbreak
(A)^{-1})\allowbreak (-\frac{3}{256}(B)^{4}\allowbreak
(A)^{-4}+\allowbreak
\frac{1}{16}(C)\allowbreak (B)^{2}\allowbreak (A)^{-3}-\allowbreak \frac{1}{4%
}(B)\allowbreak (D)\allowbreak (A)^{-2}+\allowbreak (E)\allowbreak
(A)^{-1})-\allowbreak \frac{1}{8}(\frac{1}{8}(B)^{3}\allowbreak
(A)^{-3}-\allowbreak \frac{1}{2}(B)\allowbreak (C)\allowbreak
(A)^{-2}+\allowbreak (D)\allowbreak (A)^{-1})^{2})\allowbreak +(\frac{1}{4}%
\allowbreak (-\frac{1}{108}\allowbreak (-\frac{3}{8}(B)^{2}\allowbreak
(A)^{-2}+\allowbreak (C)\allowbreak (A)^{-1})^{3}+\allowbreak \frac{1}{3}(-%
\frac{3}{8}(B)^{2}\allowbreak (A)^{-2}+\allowbreak (C)\allowbreak
(A)^{-1})\allowbreak (-\frac{3}{256}(B)^{4}\allowbreak
(A)^{-4}+\allowbreak
\frac{1}{16}(C)\allowbreak (B)^{2}\allowbreak (A)^{-3}-\allowbreak \frac{1}{4%
}(B)\allowbreak (D)\allowbreak (A)^{-2}+\allowbreak (E)\allowbreak
(A)^{-1})-\allowbreak \frac{1}{8}(\frac{1}{8}(B)^{3}\allowbreak
(A)^{-3}-\allowbreak \frac{1}{2}(B)\allowbreak (C)\allowbreak
(A)^{-2}+\allowbreak (D)\allowbreak (A)^{-1})^{2})^{2}+\allowbreak \frac{1}{%
27}\allowbreak (-\frac{1}{12}\allowbreak
(-\frac{3}{8}(B)^{2}\allowbreak
(A)^{-2}+\allowbreak (C)\allowbreak (A)^{-1})^{2}-\allowbreak (-\frac{3}{256}%
(B)^{4}\allowbreak (A)^{-4}+\allowbreak \frac{1}{16}(C)\allowbreak
(B)^{2}\allowbreak (A)^{-3}-\allowbreak \frac{1}{4}(B)\allowbreak
(D)\allowbreak (A)^{-2}+\allowbreak (E)\allowbreak (A)^{-1}))^{3})^{\frac{1}{%
2}}))^{\frac{1}{3}})^{-1})-\allowbreak 2\allowbreak (\frac{1}{8}%
(B)^{3}\allowbreak (A)^{-3}-\allowbreak \frac{1}{2}(B)\allowbreak
(C)\allowbreak (A)^{-2}+\allowbreak (D)\allowbreak (A)^{-1})\allowbreak (((-%
\frac{3}{8}(B)^{2}\allowbreak (A)^{-2}+\allowbreak (C)\allowbreak
(A)^{-1})+\allowbreak 2(-\frac{5}{6}\allowbreak (-\frac{3}{8}%
(B)^{2}\allowbreak (A)^{-2}+\allowbreak (C)\allowbreak
(A)^{-1})+\allowbreak (((-\frac{1}{2}(-\frac{1}{108}\allowbreak
(-\frac{3}{8}(B)^{2}\allowbreak
(A)^{-2}+\allowbreak (C)\allowbreak (A)^{-1})^{3}+\allowbreak \frac{1}{3}(-%
\frac{3}{8}(B)^{2}\allowbreak (A)^{-2}+\allowbreak (C)\allowbreak
(A)^{-1})\allowbreak (-\frac{3}{256}(B)^{4}\allowbreak
(A)^{-4}+\allowbreak
\frac{1}{16}(C)\allowbreak (B)^{2}\allowbreak (A)^{-3}-\allowbreak \frac{1}{4%
}(B)\allowbreak (D)\allowbreak (A)^{-2}+\allowbreak (E)\allowbreak
(A)^{-1})-\allowbreak \frac{1}{8}(\frac{1}{8}(B)^{3}\allowbreak
(A)^{-3}-\allowbreak \frac{1}{2}(B)\allowbreak (C)\allowbreak
(A)^{-2}+\allowbreak (D)\allowbreak (A)^{-1})^{2})\allowbreak +(\frac{1}{4}%
\allowbreak (-\frac{1}{108}\allowbreak (-\frac{3}{8}(B)^{2}\allowbreak
(A)^{-2}+\allowbreak (C)\allowbreak (A)^{-1})^{3}+\allowbreak \frac{1}{3}(-%
\frac{3}{8}(B)^{2}\allowbreak (A)^{-2}+\allowbreak (C)\allowbreak
(A)^{-1})\allowbreak (-\frac{3}{256}(B)^{4}\allowbreak
(A)^{-4}+\allowbreak
\frac{1}{16}(C)\allowbreak (B)^{2}\allowbreak (A)^{-3}-\allowbreak \frac{1}{4%
}(B)\allowbreak (D)\allowbreak (A)^{-2}+\allowbreak (E)\allowbreak
(A)^{-1})-\allowbreak \frac{1}{8}(\frac{1}{8}(B)^{3}\allowbreak
(A)^{-3}-\allowbreak \frac{1}{2}(B)\allowbreak (C)\allowbreak
(A)^{-2}+\allowbreak (D)\allowbreak (A)^{-1})^{2})^{2}+\allowbreak \frac{1}{%
27}\allowbreak (-\frac{1}{12}\allowbreak
(-\frac{3}{8}(B)^{2}\allowbreak
(A)^{-2}+\allowbreak (C)\allowbreak (A)^{-1})^{2}-\allowbreak (-\frac{3}{256}%
(B)^{4}\allowbreak (A)^{-4}+\allowbreak \frac{1}{16}(C)\allowbreak
(B)^{2}\allowbreak (A)^{-3}-\allowbreak \frac{1}{4}(B)\allowbreak
(D)\allowbreak (A)^{-2}+\allowbreak (E)\allowbreak (A)^{-1}))^{3})^{\frac{1}{%
2}}))^{\frac{1}{3}})-\allowbreak \frac{1}{3}(-\frac{1}{12}\allowbreak (-%
\frac{3}{8}(B)^{2}\allowbreak (A)^{-2}+\allowbreak (C)\allowbreak
(A)^{-1})^{2}-\allowbreak (-\frac{3}{256}(B)^{4}\allowbreak
(A)^{-4}+\allowbreak \frac{1}{16}(C)\allowbreak (B)^{2}\allowbreak
(A)^{-3}-\allowbreak \frac{1}{4}(B)\allowbreak (D)\allowbreak
(A)^{-2}+\allowbreak (E)\allowbreak (A)^{-1}))\allowbreak (((-\frac{1}{2}(-%
\frac{1}{108}\allowbreak (-\frac{3}{8}(B)^{2}\allowbreak
(A)^{-2}+\allowbreak (C)\allowbreak (A)^{-1})^{3}+\allowbreak \frac{1}{3}(-%
\frac{3}{8}(B)^{2}\allowbreak (A)^{-2}+\allowbreak (C)\allowbreak
(A)^{-1})\allowbreak (-\frac{3}{256}(B)^{4}\allowbreak
(A)^{-4}+\allowbreak
\frac{1}{16}(C)\allowbreak (B)^{2}\allowbreak (A)^{-3}-\allowbreak \frac{1}{4%
}(B)\allowbreak (D)\allowbreak (A)^{-2}+\allowbreak (E)\allowbreak
(A)^{-1})-\allowbreak \frac{1}{8}(\frac{1}{8}(B)^{3}\allowbreak
(A)^{-3}-\allowbreak \frac{1}{2}(B)\allowbreak (C)\allowbreak
(A)^{-2}+\allowbreak (D)\allowbreak (A)^{-1})^{2})\allowbreak +(\frac{1}{4}%
\allowbreak (-\frac{1}{108}\allowbreak (-\frac{3}{8}(B)^{2}\allowbreak
(A)^{-2}+\allowbreak (C)\allowbreak (A)^{-1})^{3}+\allowbreak \frac{1}{3}(-%
\frac{3}{8}(B)^{2}\allowbreak (A)^{-2}+\allowbreak (C)\allowbreak
(A)^{-1})\allowbreak (-\frac{3}{256}(B)^{4}\allowbreak
(A)^{-4}+\allowbreak
\frac{1}{16}(C)\allowbreak (B)^{2}\allowbreak (A)^{-3}-\allowbreak \frac{1}{4%
}(B)\allowbreak (D)\allowbreak (A)^{-2}+\allowbreak (E)\allowbreak
(A)^{-1})-\allowbreak \frac{1}{8}(\frac{1}{8}(B)^{3}\allowbreak
(A)^{-3}-\allowbreak \frac{1}{2}(B)\allowbreak (C)\allowbreak
(A)^{-2}+\allowbreak (D)\allowbreak (A)^{-1})^{2})^{2}+\allowbreak \frac{1}{%
27}\allowbreak (-\frac{1}{12}\allowbreak
(-\frac{3}{8}(B)^{2}\allowbreak
(A)^{-2}+\allowbreak (C)\allowbreak (A)^{-1})^{2}-\allowbreak (-\frac{3}{256}%
(B)^{4}\allowbreak (A)^{-4}+\allowbreak \frac{1}{16}(C)\allowbreak
(B)^{2}\allowbreak (A)^{-3}-\allowbreak \frac{1}{4}(B)\allowbreak
(D)\allowbreak (A)^{-2}+\allowbreak (E)\allowbreak (A)^{-1}))^{3})^{\frac{1}{%
2}}))^{\frac{1}{3}})^{-1}))^{\frac{1}{2}})^{-1}))^{\frac{1}{2}}).$

\begin{remark}
The construction of the roots is done by a text editor, replacing the
parameters ($A, B, C, D,$ and $E$),of the previous root formulae with
the coefficients of given polynomials. The parenthesis are necessary to
avoid errors when complex term are replaced, by example, replacing
$-256Gx^{4}+1024Gx^{6}-576x^{2}-48 - 64Gx^{2}+2816x^{4}+16G+1024x^{6}$
by $D$ for the formulae of Ferrari.
\end{remark}

\bibliographystyle{abbrv}

\end{document}